\documentclass[twocolumn,twocolappendix,numbers,sort,compress]{aastex631}

\usepackage[normalem]{ulem}

\usepackage{bm}
\usepackage{multirow} 

\usepackage{enumitem}
\usepackage[utf8]{inputenc}
\usepackage{textcomp}
\usepackage{booktabs} 
\usepackage{xcolor}

\graphicspath{{figures/}}

\begin{document}

\title{From filaments to clumps: filament properties with synthetic Herschel observations}

\author{Zhen-Xing Ma}
\affiliation{Guangxi Key Laboratory for Relativistic Astrophysics, Department of Physics, Guangxi University, Nanning 530004, China}

\author[0000-0002-9138-5940]{Zu-Jia Lu}\thanks{luzujia@gxu.edu.cn}
\affiliation{Guangxi Key Laboratory for Relativistic Astrophysics, Department of Physics, Guangxi University, Nanning 530004, China}
\affiliation{Department of Physics, University of Oxford, Keble Road, Oxford OX1 3RH, UK}

\begin{abstract}

Systematic surveys of filaments have been conducted to study their properties and their relationship to the process of star formation. In this paper, we use synthetic Herschel observations derived from 3D numerical simulations to compute column density maps, then use the \texttt{FILFINDER} algorithm to identify filaments. We obtain a large sample of 8,832 filaments that we further decompose into 110,193 branches. We characterize the physical properties of these filamentary structures and explore their correlations with embedded clumps. Furthermore, we directly compare our synthetic results with an observational catalogue of 32,059 filaments from the Herschel Infrared Galactic Plane Survey (Hi-GAL). Our results show that filaments are central to the star formation process, hosting $94\%$ of clumps from synthetic observations and $93\%$ of stars from our 3D numerical simulation. Filaments that host clumps have higher median column densities ($1.1\times10^{21}\,\rm{cm}^{-2}$) than those without ($3.8\times10^{20}\,\rm{cm}^{-2}$). We find power-law distributions for our synthetic filament masses and lengths, with power-law indexes of $\alpha_{\rm M}=-0.86$ and $\alpha_{\rm L} = -1.71$, respectively. We also find that the relation between the density of filaments and the background density is $N_{\rm{fs}} \propto N_{\rm{bs}}^{0.78}$. The measured properties of the filaments from the 2D synthetic observations are qualitatively consistent with those of the filaments from the Hi-GAL survey.

\end{abstract}


\section{Introduction}
\label{Introduction}

The formation of stars, particularly massive stars ($M \gtrsim 8 \rm\,M_\odot$), is one of the most fundamental processes driving the evolution of galaxies. Massive stars exert profound effects on their surroundings through radiation, stellar winds, and supernova explosions \citep{Zinnecker+2007ARA&A, Motte+2018ARA&A,Beuther+2025ARA&A}, thereby regulating the efficiency of star formation \citep{Geen+2018MNRAS, Potdar+2022MNRAS, Fichtner+2024A&A} and the dynamics of the interstellar medium (ISM) \citep{Ibanez+MacLow+2017ApJ, Seifried+2018ApJ, Lu+2020ApJ}. Historically, two mainstream theoretical models have been proposed to explain how these stars assemble within clumps, ranging from core accretion \citep{McKee+2002Natur} to competitive accretion \citep{Bonnell+2006MNRAS}. In the modern paradigm, however, filamentary structures are recognized as the fundamental building blocks of molecular clouds. Extensive observations, especially those from the Herschel Space Observatory, have revealed that these pervasive structures play a central role in the assembly of stellar mass by channeling gas flows and concentrating dense material into star-forming clumps or cores\citep{Andre.P+2010A&A, Molinari+2011A&A, Andre+2014prpl}.

Toward smaller scales, observational and theoretical studies increasingly suggest that filamentary structures are essential for channeling the gas flows required to form massive stars \citep{Zernickel+2013A&A, Trevio-Morales2019A&A, Kumar+2022A&A}. Observations of Infrared Dark Clouds (IRDCs), for example, suggest that filaments act as precursors to star clusters \citep{Bergin+2007ARA&A, Chambers+2009ApJS}. In particular, hub-filament systems (HFSs), structurally defined by the convergence of three or more filaments onto a central dense hub \citep{Myers+2009ApJ}, have been identified as efficient environments for sustaining the high accretion rates required to form massive stars \citep[e.g.,][]{Schneider+2010A&A, Schneider+2012A&A, Peretto+2013A&A, Peretto+2014A&A, Henshaw+2014MNRAS, Zhang+2015ApJ, Lu+2018ApJ, Xu+2023MNRAS}. While the ubiquity of this mode is still debated, new theoretical scenarios, such as the global hierarchical collapse model \citep{Vzquez-Semadeni+2019MNRAS} and the inertial-inflow model \citep{Padoan+2020ApJ}, highlight the importance of these multi-scale accretion processes. Recent surveys have begun to statistically characterize these structures. For instance, \citet{Soler+2022A&A} linked filament orientation to Galactic dynamics, while \citet{Ge+2023A&A} and \citet{Rawat+2024MNRAS} identified filaments that appear to feed central clumps.

Interpreting these observational results remains challenging due to inherent limitations. Real observations provide only two-dimensional (2D) projected views of a complex, turbulent three-dimensional (3D) environment. Projection effects, blending, and resolution limits, especially at the large distances typical of massive star-forming regions, can significantly bias the derivation of key physical quantities, such as filament width, length, and the line mass function \citep{Schisano+2014ApJ,Feng+2024MN}, and the properties of clumps and prestellar cores \citep{Lu-Zu-Jia+2022MNRAS,Padoan+2023}. Consequently, it is difficult to determine from observations alone whether the derived statistical properties accurately reflect the underlying 3D physics of the clouds.

In this work, we investigate the properties of the filaments generated from a numerical simulation and compare them with those of filaments from large-scale surveys. Our main goal is to use synthetic observations to interpret real observations on the role of filaments in star formation. We first compute the column density from synthetic Herschel observations \citep{Lu-Zu-Jia+2022MNRAS}, then use the \texttt{FILFINDER} algorithm to identify filaments \citep{Koch+2015MNRAS}. We obtain a large sample of 8,832 filaments to compare with the catalogue of 32,059 Galactic filaments from the Herschel Infrared Galactic Plane Survey (Hi-GAL) presented in \citet{Schisano+2020MN}.

The structure of the paper is as follows. In Section~\ref{Data_and_Methods} we summarize the synthetic observations and the methods to identify filaments and clumps. In Section~\ref{Results} we present the physical properties of the identified filaments, their scaling relations, and their association with clumps and stars. The implications of our findings for the stability and evolution of star-forming clouds are discussed in Section~\ref{Discussion}. Finally, the main conclusions are summarized in Section~\ref{Conclusion}.

\section{Data and Methods}
\label{Data_and_Methods}

\subsection{3D MHD simulation}
\label{simulaiton}

This work is based on the large-scale 3D magnetohydrodynamic (MHD) star-formation simulation that describes an interstellar medium (ISM) region of size $L_{\rm box}=250$ pc with a highest resolution of 0.008~pc and total mass $1.9\times 10^6\,M_{\odot}$, where the turbulence is driven by supernovae (SNe) alone \citep{Padoan+SN1+2016ApJ}. The 3D MHD simulation was carried out with the AMR code RAMSES \citep{Teyssier+2002A&A}, with periodic boundary conditions. The simulation was initially started with zero velocity, a uniform density, $n_{\rm H,0}=5$ cm$^{-3}$, a uniform temperature, $T_0=10^4$ K, and a uniform magnetic field $B_0=4.6$ $\mu$G. SN explosions were randomly distributed in space and time, with a rate of 6.25 SNe Myr$^{-1}$. This initial phase without gravity was described in \citet{Padoan+SN1+2016ApJ}. At the time $t=55.5$ Myr from the beginning of the simulation, gravity was introduced and followed the star-formation process and the resulting SNe self-consistently. The simulation was continued running for an additional period of approximately 30~Myr with gravity. This second phase of the simulation was described in \citet{Padoan+17SN_IV}. The spatial resolution of the simulation is sufficient to resolve the formation of individual massive stars \citep{Padoan+2020ApJ}. To follow the collapse of prestellar cores, sink particles were created in cells where the gas density was larger than $10^6$ cm$^{-3}$, if (i) the gravitational potential has a local minimum value, (ii) the three-dimensional velocity divergence was negative, and (iii) no other previously created sink particle was present within an exclusion radius $r_{\rm excl} = 0.12$~pc.

The ``real'' SNe are the outcome of the natural evolution of individual massive stars, which form in the period after the gravity involved (at $t=55.5$ Myr). So the SNe are spatially resolved in the simulation, meaning their position and timing are self-consistent, rather than being added randomly in space and time. The star formation, the subsequent evolution of massive stars to SN explosions, and the turbulence driven by those SNe in the 3D simulation box are all consistent with each other. For the details of the simulation, we refer the reader to previous publications \citep{Pan+16SN_II,Padoan+16SN_III,Padoan+SN1+2016ApJ,Padoan+17SN_IV,Padoan+2020ApJ}. 

\subsection{Synthetic observations and clumps}
\label{synthetic_observation}

The synthetic observations are computed in the Herschel bands (70, 160, 250, 350, and 500~\micron) in three orthogonal directions of three simulation snapshots, at assumed distances of 2, 4, 8, and 12~kpc. The snapshots correspond to 15.4, 23.3, and 34.2~$\rm{Myr}$ after the onset of self-gravity and star formation in the simulation. The calculations were carried out with the radiative transfer code SOC \citep{Juvela_2019}, using the full spatial resolution of the simulation. Details of the synthetic observation pipeline are given in \citet{Lu-Zu-Jia+2022MNRAS}.

\begin{figure}
\centering  
\includegraphics[width=1.0\linewidth]{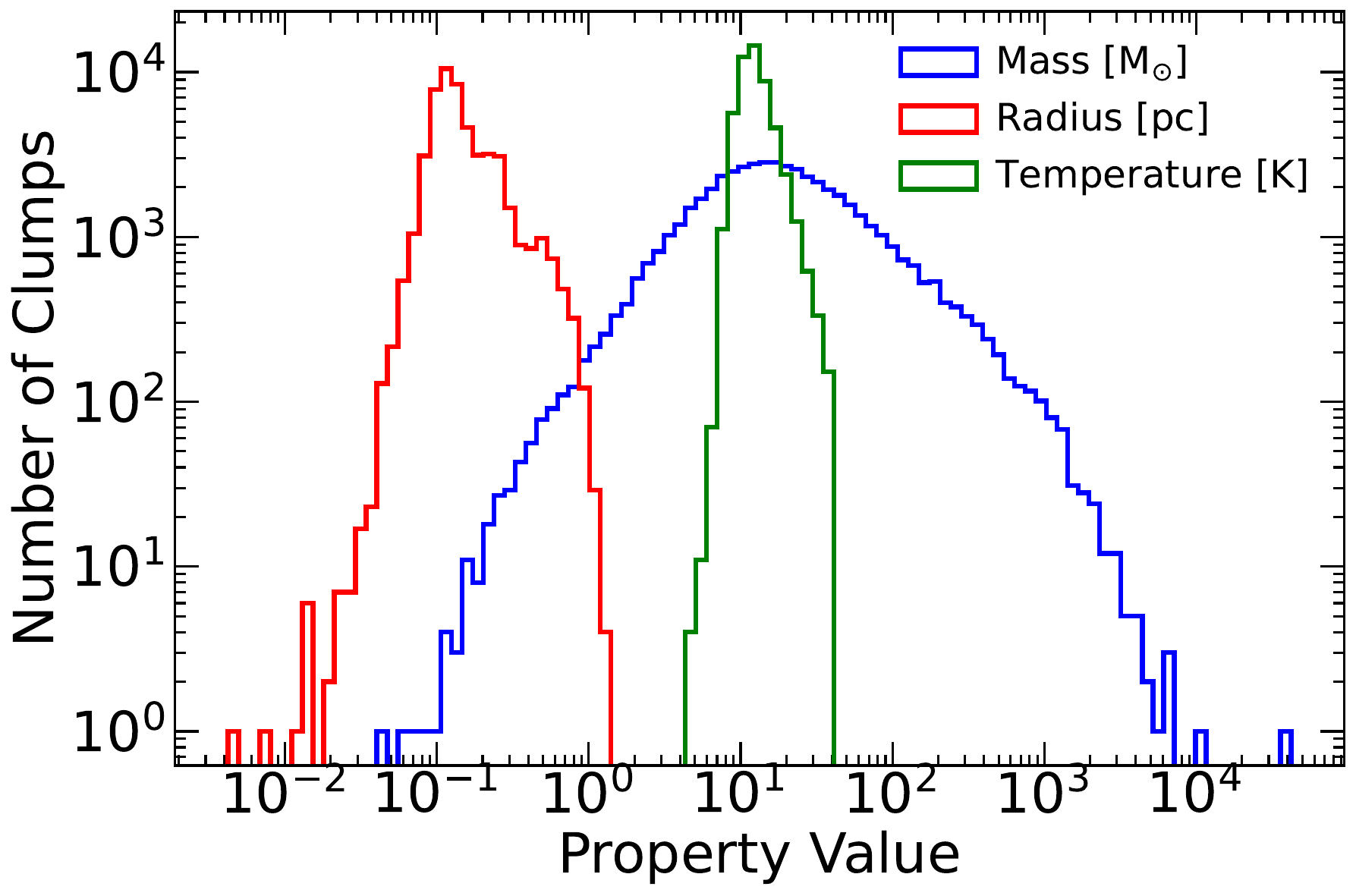}
\caption{The distributions of physical properties of 51,831 clumps used in this work, which were presented in \citet{Lu-Zu-Jia+2022MNRAS}. The blue, red and green lines are the masses of the clumps derived from SED $M_{\rm SED}$, the effective radius $R$, and the temperature of the clumps $T$, respectively. }
\label{fig_1}
\end{figure}

\begin{figure*}
\centering  
\includegraphics[width=1.0\linewidth]{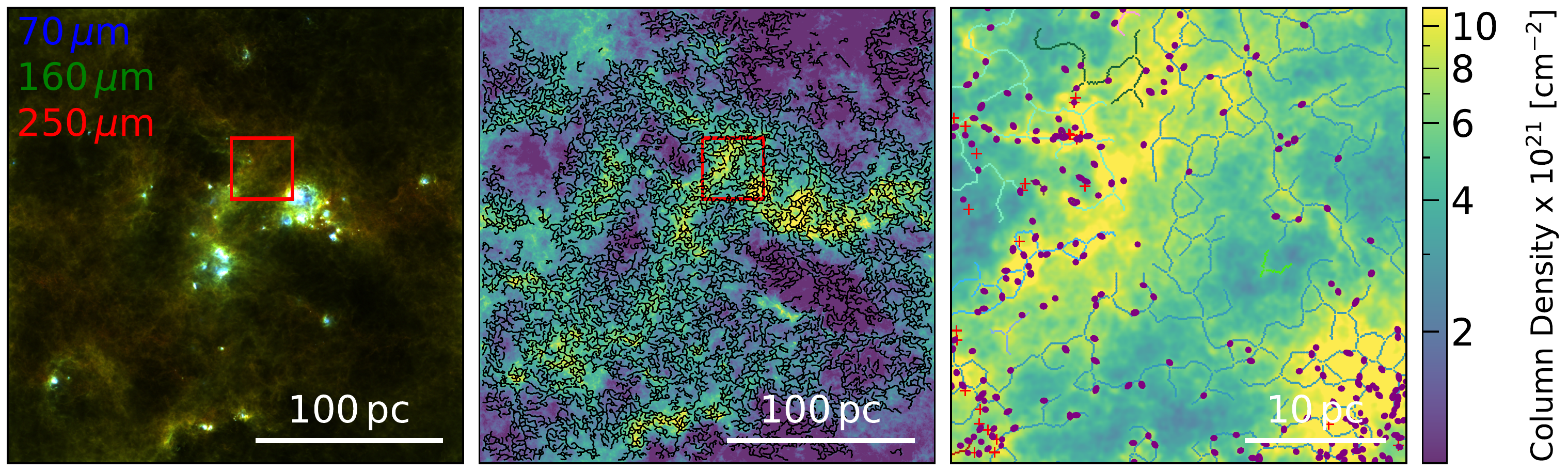}
\caption{{\it Left panel:} synthetic Herschel three-color image of the whole 250~pc simulation at an assumed distance of 2~kpc, with blue, green and red for 70, 160 and 250~\micron, respectively. \textit{Middle panel:} column density derived from the synthetic Herschel observation as shown in the left panel. The black lines show the filamentary structures. \textit{Right panel:} a zoom-in region marked in the middle panel. The purple ellipses indicate clumps and red crosses mark the position of stars. The different colors of the skeleton lines are used to distinguish individual extracted filaments.} 
\label{fig_2}
\end{figure*}

Clumps were identified from these synthetic observations in \citet{Lu-Zu-Jia+2022MNRAS}, using the \texttt{CuTEx} algorithm \citep{Molinari+2011A&A}. A compact source catalogue was compiled following the exact same data-reduction pipeline as for the Hi-GAL compact source catalogue \citep{Elia+2017, Elia+2021MNRAS}. The ``clumps'' in this work refer to the compact sources that belong to the common area of PACS+SPIRE, which are two of the three instruments aboard Herschel, and were detected at least in three consecutive Herschel bands, as defined in the Hi-GAL compact source catalogue in \citet{Elia+2017, Elia+2021MNRAS}. The clumps associated with the sample of filaments in \citet{Schisano+2020MN}, with which we compare, are from the same Hi-GAL compact source catalogue. ``2D clump" and ``3D clump" refer to the synthetic clumps detected from synthetic surface brightness maps and the corresponding main clumps identified along the line-of-sight in the column of 3D simulation, respectively. Figure~\ref{fig_1} shows the main physical properties of the 51,831 clumps in our synthetic catalogue. In this work, to investigate the role of filamentary structures in star-forming regions, we also compute column-density maps from the synthetic surface brightness, which we then use to identify filaments. 

\subsection{Column-density and temperature maps from the synthetic observations}
\label{Herschel_Column_Density_Maps}

We compute column density $N(\mathrm{H}_2)$ and temperature maps from our synthetic observations following the approach described in \citet{Elia+2013ApJ...772...45E, Schisano+2014ApJ, Schisano+2020MN}. We use the four Herschel bands at the wavelengths 160, 250, 350, and 500~\micron\ that capture emission from cold grains, while the 70~\micron~band was excluded due to contamination from warmer dust components, such as circumstellar material and H\textsc{ii} regions \citep[e.g.,][]{Schneider+2012A&A,Elia+2013ApJ...772...45E}.
~Following the same approach described in \citep{Elia+2013ApJ...772...45E, Schisano+2014ApJ, Schisano+2020MN}, all the synthetic maps at shorter wavelengths (70-350~$\mu$m) are first smoothed to a common angular resolution of $34.5^{\prime\prime}$ to match the beam size of the 500~$\mu$m band. Subsequently, the smoothed maps are regridded to a common grid with a pixel scale of $11.5^{\prime\prime}$. This pixel scale corresponds to a physical pixel size of approximately $0.11\rm\,pc$, $0.22\rm\,pc$, $0.44\rm\,pc$, and $0.67\rm\,pc$ at assumed distances of $2\rm\,kpc$, $4\rm\,kpc$, $8\rm\,kpc$, and $12\rm\,kpc$, respectively. 

We perform a single-temperature modified greybody function to fit the $\lambda$ $\geq$ 160~$\mu$m spectral energy distributions (SEDs) of the synthetic observation maps pixel by pixel to estimate $N(\mathrm{H}_2)$ and the temperature $T$, 
\begin{equation}
{F_\nu} = N(\mathrm{H}_2) \, {\mu \, m_{\mathrm{H}} \, \Delta\theta_{500}^2 \, \kappa_0 \, \left(\frac{\nu}{\nu_0}\right)^\beta \, B_\nu(T)} \,\, ,
\end{equation}
where $F_{\nu}$ is the flux density of each pixel at the frequency $\nu$, $\mu$ is the mean molecular weight assumed to be equal to 2.8 for the classical cosmic abundance ratio that accounts for a $25\%$ helium mass fraction \citep{Kauffmann+2008A&A}, $m_{\rm{H}}$ is the mass of a hydrogen atom, $\Delta\theta_{500}$ is the angular pixel size at 500~\micron\ in sr, $\kappa_{0}=0.1\,\rm{cm}^2\,\rm{g}^{-1}$ is the dust opacity at the reference frequency $\nu_{0}=1,200\,\rm{GHz}$ (equivalent to $\lambda_0=250\,\mu\rm{m}$), $\tau_{\nu}=\left( \nu /\nu_0 \right)^{\beta}$ is the optical depth, and $B_\nu(T)$ is the Planck function at the dust temperature $T$. The dust opacity spectral index $\beta$ is the exponent of the power-law dust emissivity at large wavelengths. While \citet{Hildebrand+1983QJRAS}, \citet{Schisano+2014ApJ} and \citet{Elia+2017} adopted the value of $\beta=2$, we use $\beta=1.8$, which is more appropriate for the dust model of \cite{OH_1994} that we have used in the radiative transfer calculations \citep{Lu-Zu-Jia+2022MNRAS}.

For the surface brightness relative noise values in our synthetic observations in \citet{Lu-Zu-Jia+2022MNRAS}, we assumed $4\%$ in the PACS bands ($70$ and $160\,\mu\rm{m}$), and $2\%$ in the SPIRE bands ($250$, $350$, and $500\,\mu\rm{m}$). These noise levels are consistent with actual Herschel Hi-GAL observations. To derive more accurate column densities $N(\mathrm{H}_2)$ and temperatures, we assign identical weights, which we assumed in the synthetic surface brightness maps, in SED fitting procedure as \citet{Konyves+2015A&A} did with Herschel Gould Belt survey observations. Since statistical weights scale inversely with variance ($w \propto 1/\sigma^2$), this naturally assigns higher weights to the long-wavelength bands ($\ge 250\,\mu\rm{m}$), ensuring the fit is predominantly constrained by these robust cold dust tracers. Furthermore, using fixed relative uncertainties instead of variable instrumental noise maps prevents overfitting and effectively mitigates noise fluctuations.

In Figure~\ref{fig_2}, we show an RGB image of one of the synthetic maps (left panel), and the corresponding column density including the identified filaments (middle panel). In the right panel, we show a zoom-in of the column density region marked in the previous panels, along with the clumps identified from the surface brightness maps and the stars from the 3D simulation.

\subsection{Filament detection from column density maps}
\label{identify_filaments}

The filament identification algorithm was run on 36 column-density maps corresponding to three directions of three simulation snapshots assumed to be at four different distances. The lengths of filamentary structures in the ISM range from $\sim 0.01\,\rm{pc}$ to several kiloparsecs \citep{Hacar+2023ASPC}. We identify filaments using the \texttt{FILFINDER} algorithm \citep{Koch+2015MNRAS}. To reliably separate genuine filaments from noise artifacts, we define our extraction criteria based on the instrumental resolution, where one beam (FWHM) corresponds to 3 pixels in our maps. To optimize the extraction and ensure physical validity, we systematically set the key parameters of the filament extraction algorithm based on this specific beam scale.

\begin{figure*}
\centering  
\includegraphics[width=1.0\linewidth]{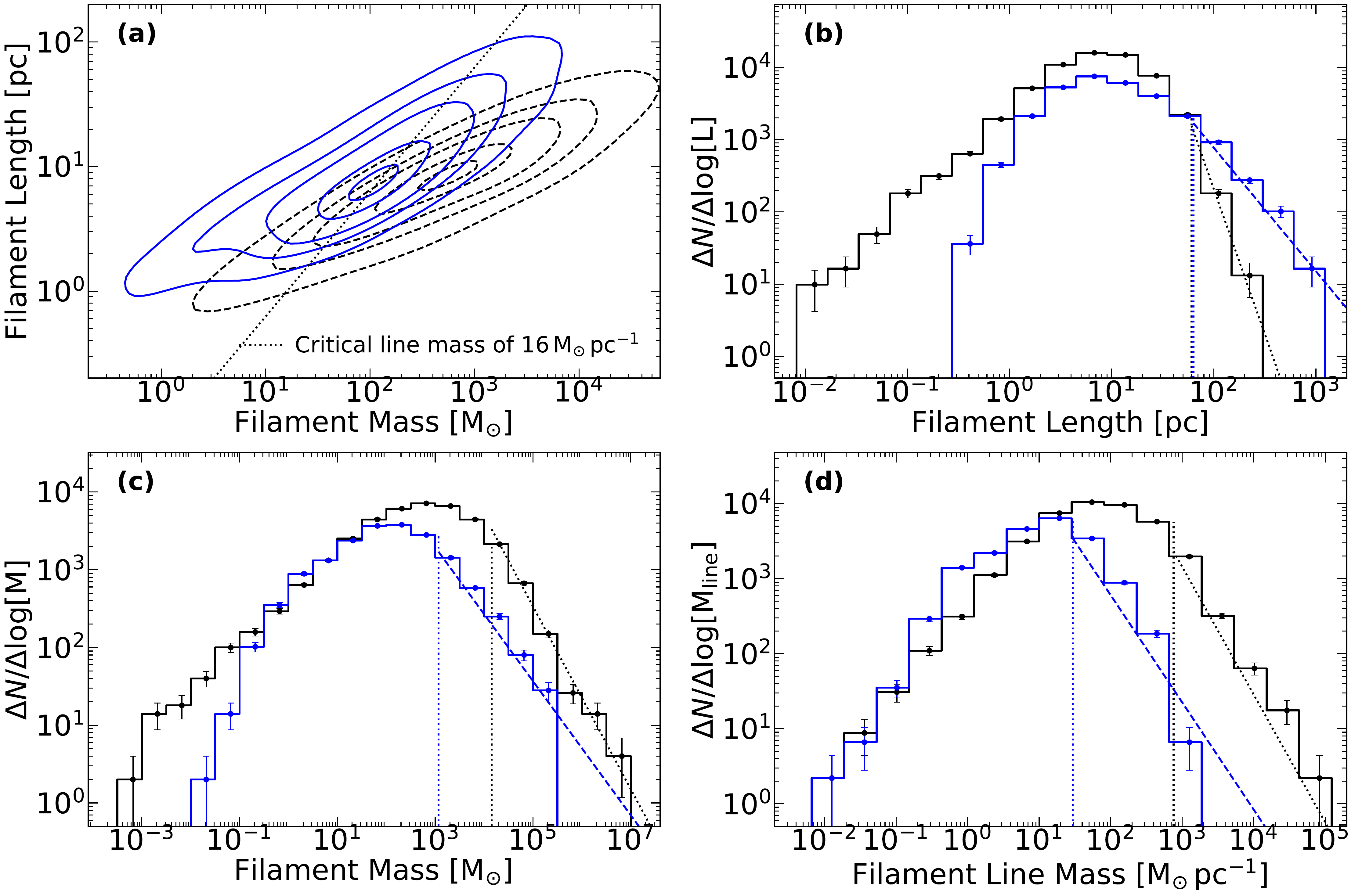}
\caption{Physical properties of the identified filaments compared with Hi-GAL filaments from \citealt{Schisano+2020MN}. $Panel\,(a)$ is the relation of branch length vs.\ mass. The contours for the synthetic (blue) and Hi-GAL (black) filaments show the number density of $90\%$, $70\%$, $50\%$, $20\%$, and $5\%$ of the maximum. The dotted line is the critical line mass of $16\,\rm M_\odot\,pc^{-1}$ at $T = 10\rm\,K$. $Panel\,(b),\,(c),\,(d)$ are the distributions of length, mass, and line mass, respectively. The masses and line masses are background-subtracted. Solid histograms show the data with Poissonian error bars ($\sqrt{N}$), and blue dashed (synthetic) and black dotted (Hi-GAL) lines indicate the power-law fits derived using the MLE method. The vertical dotted lines mark the lower limits ($x_{\min}$) adopted for the fitting of our filaments (blue) and the Hi-GAL filaments (black). The fitted slopes for our filaments (blue) and Hi-GAL (black) are: (b) $\alpha_{\mathrm{L}} = -1.71 \pm 0.07$ vs.\ $-4.02 \pm 0.34$; (c) $\alpha_{\mathrm{M}} = -0.86 \pm 0.03$ vs.\ $-1.17 \pm 0.04$; (d) $\alpha_{\mathrm{M}_{\rm line}} = -1.43 \pm 0.03$ vs.\ $-1.70 \pm 0.06$. 
}
\label{fig_3}
\end{figure*}

We first apply a spatial smoothing filter setting \texttt{smooth\_size} to 1~beam. This step suppresses sub-beam pixel-to-pixel noise while preserving resolved structural details. The global column density threshold \texttt{glob\_thresh} is set to $8 \times 10^{19}\,\mathrm{cm}^{-2}$, the minimum variations in column density of Hi-GAL maps presented in Figure 2 of \citet{Schisano+2020MN}, which effectively separates the physically significant cloud structures from the diffuse ambient background and lower-density noise regions. For local background evaluation within these masked regions, we configure the adaptive thresholding window \texttt{adapt\_thresh} to 3~beams. Chosen to be three times the typical filament width ($\sim$1~beam), this scale successfully decouples elongated structures from their immediate surroundings. To filter out compact clumps, the minimum area threshold \texttt{size\_thresh} is established at 45 $\rm{pixels}^2$. This physically corresponds to the area of a minimal filament modeled as an ellipse with a width of 1 beam and a length of 5~beams \citep{Koch+2015MNRAS}. Finally, during the skeletonization and pruning phase, the minimum skeleton length \texttt{skel\_thresh} is required to be 5~beams to ensure a sufficiently elongated aspect ratio, and the minimum branch length \texttt{branch\_thresh} is 3~beams to remove short, artificial spurs caused by boundary noise \citep{Koch+2015MNRAS}. The rest of the parameters are set to their default values.

Following the \texttt{FILFINDER} run with the parameters described above, we then do additional selection on its output with the criteria that require a minimum aspect ratio of 3 and a filament-to-background contrast ratio greater than 0.3 as in \citet{Arzoumanian+2019A&A}. The above procedure yields 8,832 \emph{filaments} from the 36 column-density maps. These structures are further decomposed into 110,193 \emph{branches} defined as the paths between two filament junctions or between a junction and an endpoint, following \citet{Schisano+2014ApJ, Schisano+2020MN}. In the right panel of Figure~\ref{fig_2}, we show a 30~pc region including all the identified filaments, the clumps (purple ellipses), and the positions of the stars in the simulation (red crosses). 

The catalogue of 32,059 Galactic filaments in \citet{Schisano+2020MN}, which we compare with our synthetic observations in this work, were detected using a Hessian-matrix filament extraction algorithm from Herschel observations. Since that specific code is not public, we instead performed the \texttt{FILFINDER} algorithm \citep{Koch+2015MNRAS} on our synthetic observations. \texttt{FILFINDER} code is a robust, open-source tool extensively benchmarked and widely used in both observations \citep[e.g.,][]{Howard+2021MN,Zhou+2022MN} and simulations \citep[e.g.,][]{Feng+2024MN,Pillsworth+2025ApJ}.

\subsection{Criteria for clumps on/off filaments}
\label{criteria_clump_on_filament}

To identify the spatial associations between clumps and filaments, we perform a 2D intersection analysis. The geometric matching procedures are implemented based on the specific structural parameters available in the synthetic and observational catalogues.

For the synthetic observations, the filaments are represented by the 1D spine coordinates extracted with the \texttt{FilFinder} algorithm. Since these spines only trace the central lines of the filaments, we account for the missing spatial extent by defining a clump's extended boundary with a radius $R_{\mathrm{out}} = 2.5 \times (\mathrm{FWHM}_x + \mathrm{FWHM}_y)$. If at least one pixel of a filament's spine falls within this boundary, the clump is classified as ``on-filament'', and the intersected filament is identified as a clump-hosting filament. Furthermore, if a clump intersects with three or more distinct branches, we consider it to be located in a ``hub'' structure.

For the Herschel Hi-GAL observation, clumps are modeled as circles with a radius $R_c = \mathrm{FWHM}_{250} / 2$. Each filament is geometrically reconstructed as a rotated rectangle, where its center is placed at the cataloged central coordinates, its length and width correspond to the cataloged semi-major and semi-minor axes, and its alignment is determined by the specific position angle. To accurately reproduce the pixel-based association statistics established by \citet{Schisano+2020MN}, we enlarge this rectangular region by applying an empirical expansion factor of $1.4$ to the filament's axes. Finally, we evaluate the minimum geometric distance from the clump's center to this expanded rectangle. If this distance is less than or equal to $R_c$, the clump is classified as ``on-filament'', and the corresponding filament is identified as a clump-hosting filament.

\section{Results}
\label{Results}

\subsection{Physical properties of the filaments}
\label{physical_properties_of_the_filament}

We begin our analysis by characterizing the fundamental physical properties of the identified filaments. In Figure~\ref{fig_3}, we present the properties of our filaments (blue lines) and filaments from the Hi-GAL survey presented in \citet{Schisano+2020MN} (black lines), including the mass-length relation and the distributions of filament mass, length and line mass.

Panel (a) of Figure~\ref{fig_3} shows the correlation between the mass of filaments and their length. The dotted line is the thermal critical line mass of $16\rm \,M_\odot\,\rm{pc}^{-1}$ at $T = 10\rm\, K$. Both samples of synthetic and Hi-GAL observations show a positive mass–length correlation and the relations are shallower than the thermal critical line mass,  while the lengths of synthetic filaments are systematically longer than those from Hi-GAL. This offset is possibly introduced by the algorithms, as \texttt{FILFINDER} links structures into extended networks while the Hessian method tends to fragment them into shorter segments. The figure also shows that the majority of massive filaments ($\gtrsim 100\,\mathrm{M}_\odot$) from both observations and simulations are gravitationally supercritical, lying below the thermal critical line mass. 

\begin{deluxetable}{@{} l@{\extracolsep{\fill}}cccc @{}}
\tabletypesize{\footnotesize} 
\tablewidth{\columnwidth}
\tablehead{
\colhead{\multirow{2}{*}{Property}} & \multicolumn{2}{c}{Simulation} & \multicolumn{2}{c}{Observation} \\
\cline{2-3} \cline{4-5} 
\colhead{} & \colhead{$\alpha$} & \colhead{$x_{\min}$} & \colhead{$\alpha$} & \colhead{$x_{\min}$}
}
\startdata
Length (pc) & $-1.71 \pm 0.07$ & $63.13$ & $-4.02 \pm 0.34$ & $60.46$ \\
Mass ($\mathrm{M}_\odot$) & $-0.86 \pm 0.03$ & $1176.07$ & $-1.17 \pm 0.04$ & $1426.30$ \\
Line Mass ($\mathrm{M}_\odot\,\mathrm{pc}^{-1}$) & $-1.43 \pm 0.03$ & $29.49$ & $-1.70 \pm 0.06$ & $759.58$ \\
\enddata
\caption{MLE fitting results for the power-law tails of the distributions of lengths, masses, and line masses of the filaments. The power-law index $\alpha$ and the optimal lower bound $x_{\min}$ are automatically determined using the \texttt{powerlaw} package by minimizing the Kolmogorov-Smirnov distance.}
\label{table_mle_results}
\end{deluxetable}

In panel (b)-(d) of Figure~\ref{fig_3}, the distributions of filament lengths, masses, and line masses exhibit a turnover at the low-value end and a prominent power-law tail ($\mathrm{d}N/\mathrm{d}\log{X} \propto X^{\alpha}$) at the high-value end. The low-end turnovers primarily arise from resolution constraints and observational completeness limits, rather than reflecting intrinsic physical scales. To robustly characterize the power-law tails without binning biases, we employ the Maximum Likelihood Estimation (MLE) method via the \texttt{powerlaw} package \citep{Alstott+2014PLoSO}. For each fit, the optimal lower bound ($x_{\min}$) is determined by minimizing the Kolmogorov-Smirnov (K-S) distance. The derived MLE parameters, including $\alpha$, $x_{\min}$, and their uncertainties, are summarized in Table~\ref{table_mle_results}. 

As shown in panel (b) of Figure~\ref{fig_3}, the length distributions peak at a value just below 10~pc in both the synthetic and observational samples. However, the synthetic filaments exhibit a much shallower power law slope than in the observations ($\alpha_{\mathrm{L}} \approx -1.7$ vs.$-4.0$). Rather than showing a fundamental difference, these power law tails may reflect differences in the filament selection methods, as \texttt{FILFINDER} tends to generate much more connected and longer structures than the method in \citet{Schisano+2014ApJ}, including filaments with length exceeding the size of the computational box. 

Despite this difference in the length distributions, the line-mass distributions have similar slopes as shown in panel (d) of Figure~\ref{fig_3}. Our derived slope of $\alpha_{M_{\rm line}} = -1.43 \pm 0.03$ is comparable to both the Hi-GAL value ($\approx -1.70$) and the characteristic value of $\approx -1.5 \pm 0.1$ reported for thermally supercritical filaments in nearby molecular clouds \citep{Andre+2019A&A}. Furthermore, the line masses of our simulated filaments span a broad range from $\sim0.02$ to $2,000\,\rm{M}_\odot\,\rm{pc}^{-1}$, which is well within the wide distribution ($\sim0.02\text{--}10^5\,\rm{M}_\odot\,\rm{pc}^{-1}$) of the filaments from the Hi-GAL survey \citep{Schisano+2020MN}. In addition, the high-mass regime of our sample ($\sim200\text{--}2,000\,\rm{M}_\odot\,\rm{pc}^{-1}$) is consistent with the regime of dense filaments observed in the ATLASGAL survey that are preferentially associated with massive star formation \citep{Li-Guang-Xin+2016A&A}. 

Panel (c) of Figure~\ref{fig_3} shows the mass distribution of our synthetic and Hi-GAL filaments. We find a power-law distribution, which spans 2 orders of magnitude, starting from the peak at $\sim 10^3 \, \rm{M}_\odot$, about an order of magnitude lower than Hi-GAL filament masses with the peak at $\sim 10^4 \, \rm{M}_\odot$. The masses of Hi-GAL filaments range over 3 orders of magnitude from the peak. We fit a power law to the distributions of synthetic and Hi-GAL filament masses, resulting in the power-law index $\alpha_{\rm M}=$ $-0.86$ vs. $-1.17$. The mass distribution of our synthetic filaments is similar to Hi-GAL observations. Our result is also consistent with the filaments from large-scale MHD simulations. For example, \citet{Pillsworth+2025ApJ} extracted over 500 filaments ranging up to 10 kpc in scale with the same extraction algorithm \texttt{FILFINDER} but from the column density projection of the entire galaxy produced in the multiscale galactic MHD simulations in \citet{Zhao+2024ApJ}, rather than from the surface brightness maps. They used the data from the full simulated galactic disk, spanning 26 kpc across with spatial resolution $\sim 5.2$ pc. They also found the power-law distribution of filament masses, spanning over 5 orders of magnitude in masses ($\sim 10^3 - 10^8 \, \rm{M}_\odot$), with a power-law index $\alpha \sim 1.2$ \citep{Pillsworth+Erratum+2025ApJ}. 

In Appendix \ref{app:branch_properties}, Figure~\ref{fig_app_four_panel} reproduces Figure~\ref{fig_3} but compares the physical properties of synthetic branches with the Hi-GAL filaments.

\begin{figure}
\centering  
\includegraphics[width=\linewidth]{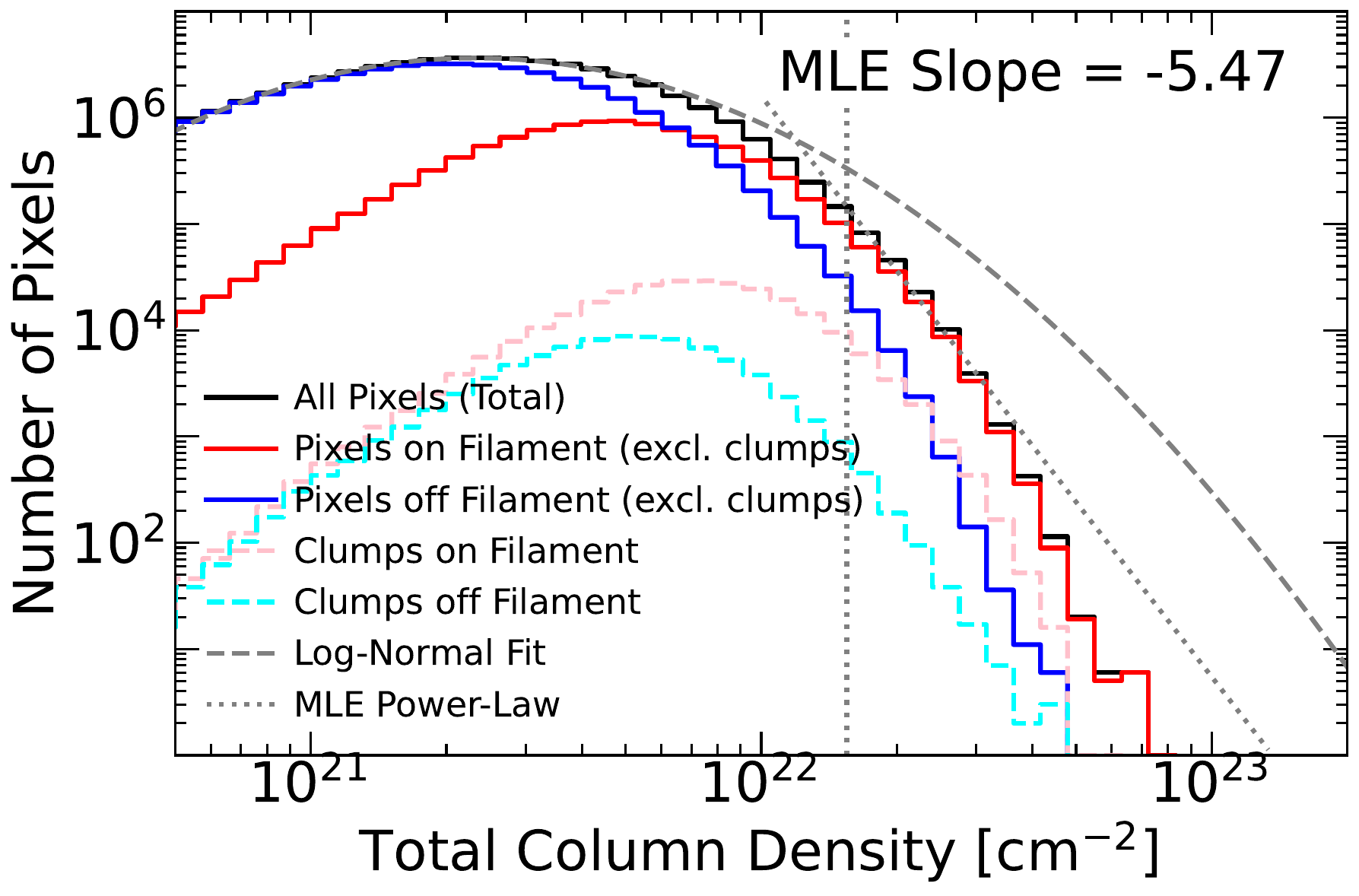} 
\caption{PDFs of map pixels. Black: all pixels; blue: non filament regions; red: filament regions; pink: clump on filament regions; cyan: clump on non filament regions. The grey dashed line represents the log-normal fit. The grey dotted line denotes the MLE power-law fit, with the slope of $-5.47$. The vertical dotted line marks the lower limit ($x_{\min}$) adopted for the power-law fitting.}
\label{fig_4}  
\end{figure}

\subsection{Probability density functions of column density}
\label{Probability_Density_Functions_of_Column_Density}

The probability density function (PDF) of column density serves as a key diagnostic of the physical processes shaping molecular clouds. A log-normal distribution of the PDF of gas density is associated with turbulent isothermal gas \citep{Vazquez-Semadeni94,Padoan+1997MNRAS} while a power-law tail at high densities indicates the onset of gravitational collapse or non-isothermal evolution \citep{Passot+1998PhRvE,Kritsuk+11}. These functional forms have been found also in the column-density PDF both in simulations \citep[e.g.][]{V'azquez-Semadeni+2001ApJ,Federrath+10} and observations \cite[e.g.][]{Kainulainen+2009A&A,Kainulainen+Tan13,Schneider+2013ApJ} of turbulent star-forming clouds. 

Here, we examine the column density PDFs of our synthetic maps to evaluate whether the density structure is dominated by turbulence or gravity and to characterize the distinct contributions from filamentary and non-filamentary regions.

Figure~\ref{fig_4} shows the PDFs of different spatial components. To better characterize the spatial contributions, we consider five categories, namely all pixels, filament regions excluding clumps, non-filament regions excluding clumps, clumps on filaments, and clumps off filaments. Non-filamentary pixels (blue) peak at $N({\mathrm{H}_2})\approx2\times10^{21}\,\rm{cm}^{-2}$, consistent with the diffuse Galactic background \citep{Schisano+2014ApJ}. Filamentary pixels (red) peak at $N(\mathrm{H}_2)\approx5\times 10^{21}\,\mathrm{cm}^{-2}$ and dominate the regions above $2\times10^{22}\,\mathrm{cm}^{-2}$. Non-filamentary pixels span a wider but lower range, from $3\times10^{18}$ to $2\times10^{21}\,\rm{cm}^{-2}$. To quantitatively evaluate the density structure, we fit the PDF peak with a log-normal distribution, and specifically use MLE to fit the power-law tail. The MLE yields a steep slope of $\alpha = - 5.47$, confirming the absence of a gravitationally driven power-law tail \citep[e.g.,][]{Schneider+2013ApJ} and simply representing the log-normal exponential cut-off. This stands in contrast to actively star-forming regions, which typically exhibit prominent power-law tails \citep[e.g.,][]{Schisano+2014ApJ, Schneider+2022A&A}. Instead, our results are consistent with the log-normal PDFs observed in diffuse, quiescent clouds where turbulence is the primary structuring agent \citep[e.g.,][]{Schneider+2015A&A,Schneider+2022A&A}. This suggests our simulated filaments are in an early, turbulence-dominated evolutionary phase prior to the onset of global collapse (see Section \ref{Limitations_and_Implications_for_synthetic_Observationals} for detailed discussion).

\subsection{Properties of column density with embedded filaments and their environment}
\label{column_density_properties_of_filaments}

\begin{figure}
\centering  
\includegraphics[width=\linewidth]{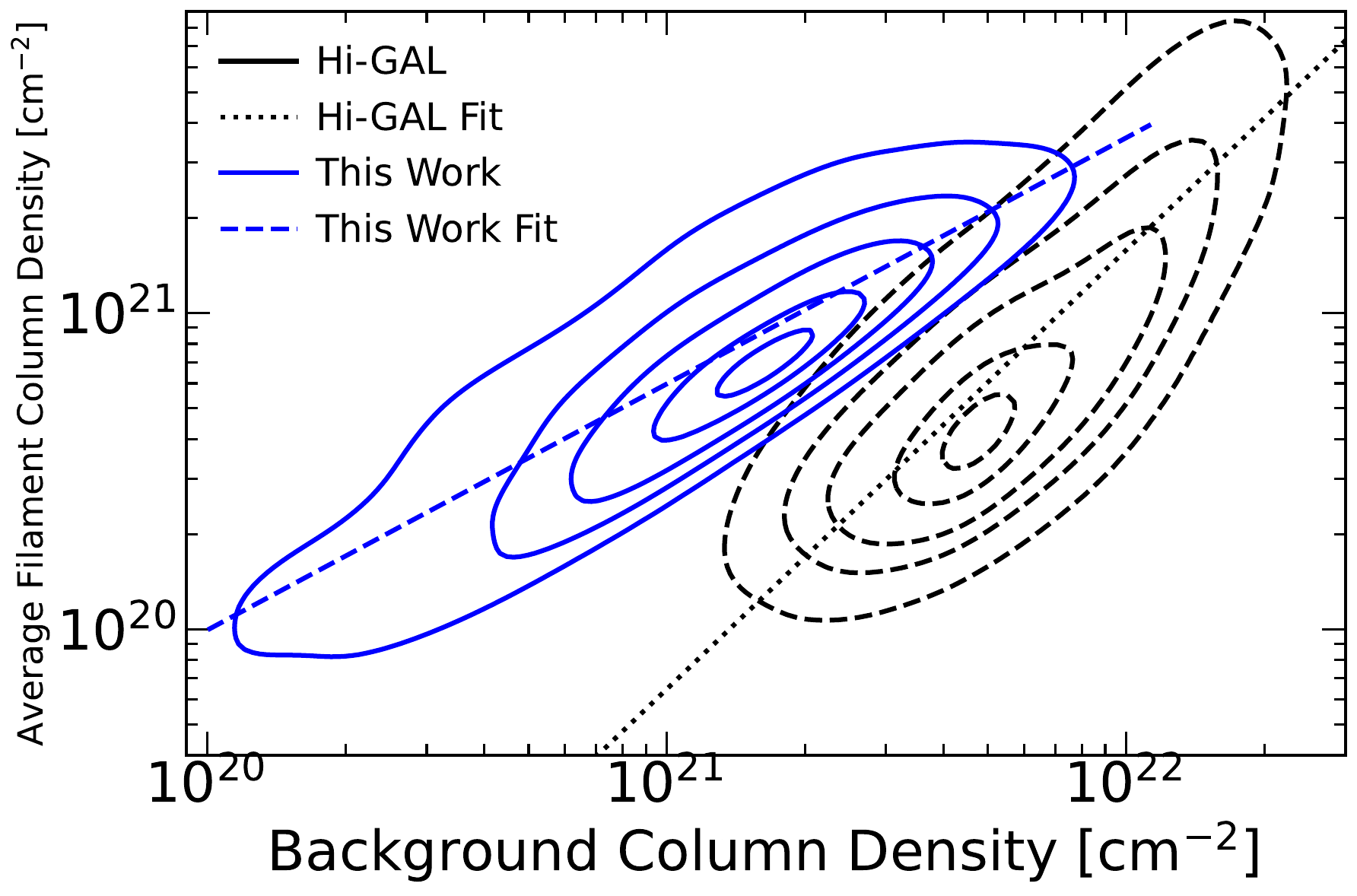} \\
\includegraphics[width=\linewidth]{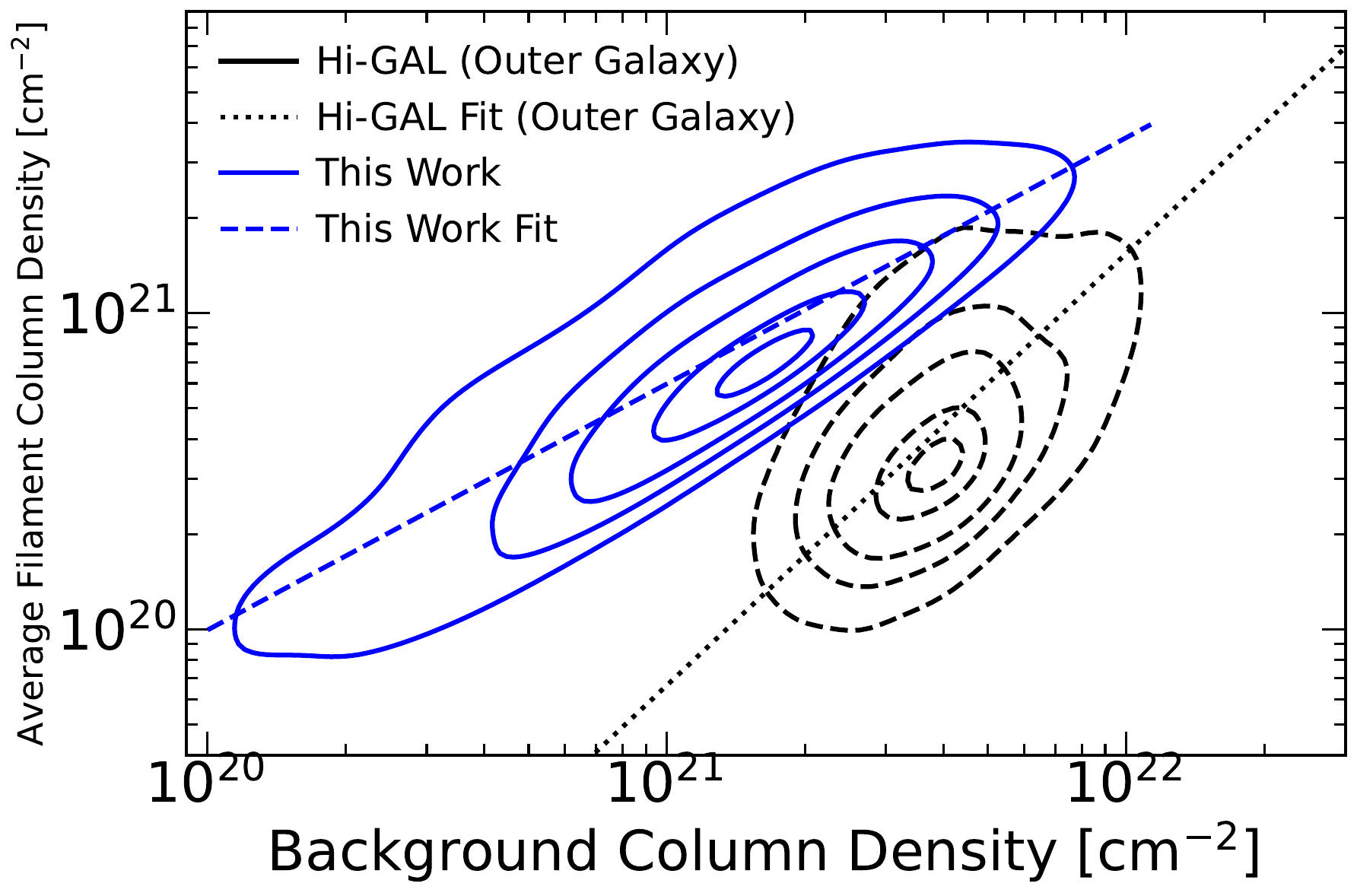}
\caption{Correlation between background column density and filament-averaged column density. \textit{Top panel}: Filament background column density versus average filament column density after subtracting the background. Contours show the point density of $90\%$, $70\%$, $50\%$, $20\%$, and $5\%$ of the maximum for synthetic observations (blue) and Hi-GAL filaments from \citet{Schisano+2020MN} (black), spanning $\rm\sim0.2-16\,kpc$. The blue dashed and black dotted lines represent the Bayesian linear regression fits to the synthetic observations and Hi-GAL filaments, respectively. \textit{Bottom panel}: the same as the top panel, but for Hi-GAL filaments only in the Outer Galaxy ($67^\circ \le l \le 289^\circ$).}
\label{fig_5}  
\end{figure}

\begin{figure}
\centering  
\includegraphics[width=\linewidth]{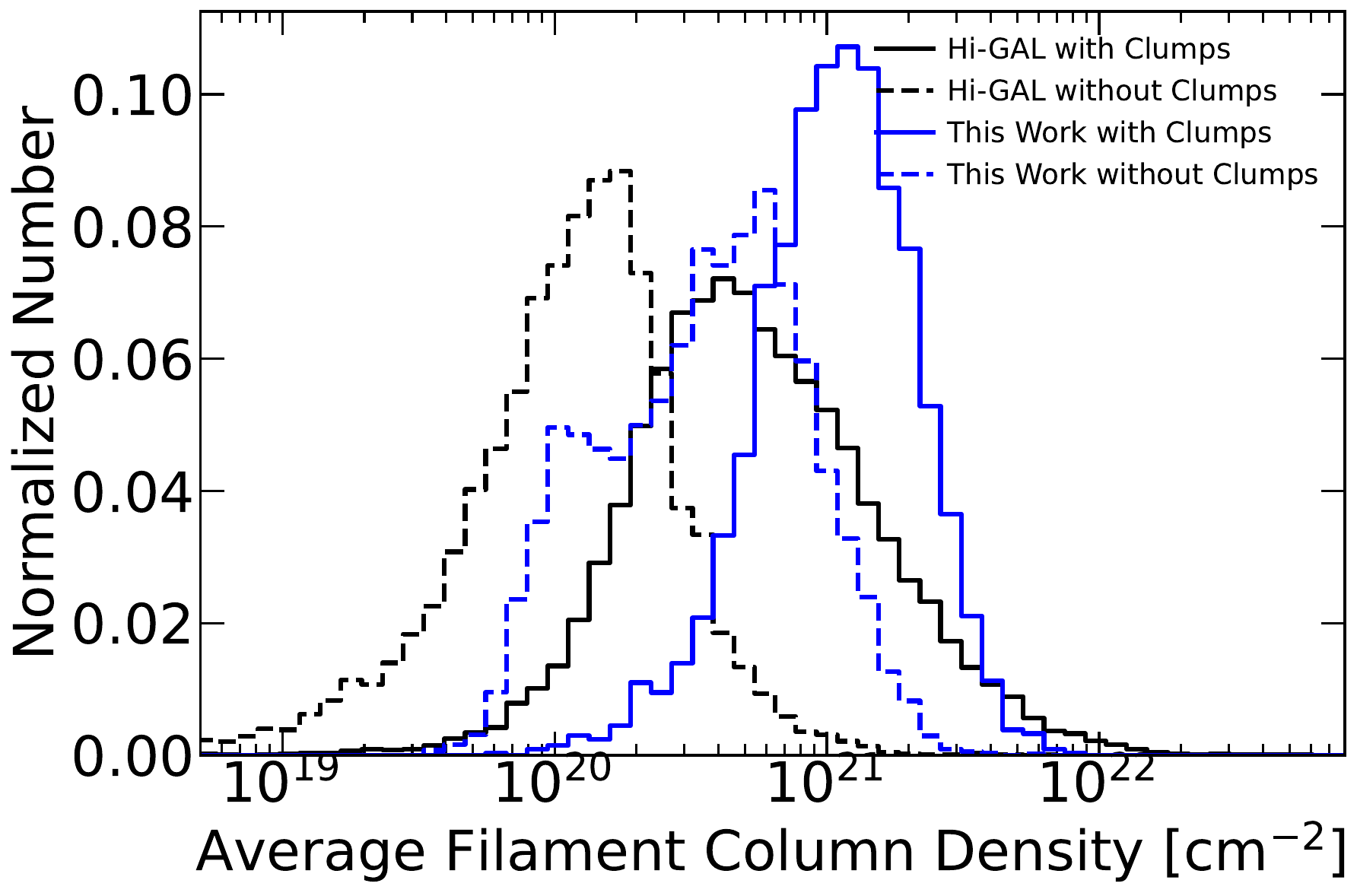}  
\caption{ Column density histograms for our synthetic filaments (blue lines) and the Hi-GAL filaments from \citet{Schisano+2020MN} (black lines). Solid lines indicate structures associated with clumps, while dashed lines indicate those without clumps.
}  
\label{fig_6}  
\end{figure}

The column densities of filaments reflect both their structural roles and the environmental conditions that prevail during star formation. We examine the correlation between the filament column density, $N_{\rm{f}}$, and the background column density, $N_{\rm b}$, where $N_{\rm b}$ is defined as the median column density of the pixels immediately surrounding the filament mask. To better trace the properties of filaments, we neglect the diffuse background with column density $N_{\rm b} < 10^{20}\,\rm{cm}^{-2}$, as this gas does not contain star-forming activity. With a Bayesian linear regression fit in the logarithmic space \citep{Kelly+2007ApJ}, the scaling relation is
\begin{equation}
N_{\rm{fs}}=(2.69\pm0.91)\times10^{4}\,N_{\rm{bs}}^{0.78\pm0.01} \,\, ,
\end{equation}
where $N_{\rm{fs}}$ and $N_{\rm{bs}}$ are the column and background density of the synthetic filament, as shown with the dashed line in Figure~\ref{fig_5}. The result shows a strong correlation between the background and average filament column densities, with a Spearman rank correlation coefficient of $\rho=0.82$ and $p<0.001$, where $\rho$ measures the strength of their monotonic relationship and $p$ indicates the probability that this correlation arises by chance. This is consistent with models of filament formation via converging flows in dense environments \citep{Schisano+2014ApJ}.

For comparison, we fit the Hi-GAL sample (black contours in the top panel of Figure~\ref{fig_5}) with the same method and obtain
\begin{equation}
N_{\rm{fo}}=(4.61\pm0.65)\times10^{-10}\,N_{\rm{bo}}^{1.39\pm0.01} \,\, , 
\end{equation}
where $N_{\rm{fo}}$ and $N_{\rm{bo}}$ are the column and background density of the Hi-GAL filament. The black dotted line fit is significantly steeper than that of our synthetic filaments, indicating a more rapid increase in filament density relative to the background. The observations have a much higher background because they integrate along the entire line-of-sight through the Galactic disk (particularly in the inner Galaxy), while our simulation only integrates only over the 250 pc extent of the computational box, which naturally results in a lower background than in Hi-GAL, as expected. This result is consistent with the findings of \citet{Lu-Zu-Jia+2022MNRAS}, which are due to line-of-sight projection effects: the Hi-GAL sources cover the inner Milky Way, so the integrated background is higher.

We divide the Hi-GAL catalogue into the Outer Galaxy (longitude range $67^{\circ}$--$289^{\circ}$) and the Inner Galaxy ($0^{\circ}$--$67^{\circ}$ and $289^{\circ}$--$360^{\circ}$). In the bottom panel of Figure~\ref{fig_5}, we plot only the Hi-GAL filaments in the Outer Galaxy for comparison. The slope of the fit for the Outer Galaxy sample ($N_{\rm{f,out}}=(2.57\pm2.09)\times10^{-9}\,N_{\rm{b,out}}^{1.36\pm0.02}$) is similar to that of the whole Hi-GAL filament sample, but the high-density part is absent, consistent with the Outer Galaxy lacking the dense, complex inner-disk regions that dominate the high-column-density tail in the full sample.

Figure~\ref{fig_6} presents the column densities of synthetic filaments with and without clumps, alongside the filament sample from the Hi-GAL survey \citep{Schisano+2020MN}. In both the Hi-GAL data (black lines) and our synthetic filaments (blue lines) there is a distinct separation between the two populations, where filaments hosting clumps exhibit significantly higher column densities than those without. However, the column density of our filaments is on average a few times higher than that of the Hi-GAL filaments. The synthetic filaments associated with clumps possess higher median column densities ($1.1\times10^{21}\,\rm{cm}^{-2}$) than those without clumps ($3.8\times10^{20}\,\rm{cm}^{-2}$). This statistically significant difference is confirmed by a two-sample Kolmogorov–Smirnov (K-S) test ($D=0.52$, $p<0.001$), where $D$ represents the maximum deviation between their cumulative distribution functions, and $p$ indicates the probability that such a difference arises by chance.

To test whether this difference could be an artifact of synthetic observational effects (e.g., resolution degradation with distance), we examined the correlation between column density and distance. A weak negative correlation was found (Spearman rank correlation test: $\rho = -0.05$, $p<0.001$), suggesting that distance does not significantly bias filament classification. Therefore, the observed differences between filaments with and without clumps probably reflect genuine density variations, rather than projection or resolution effects. The Figure~\ref{fig_app_col_dens} in Appendix \ref{app:branch_properties}, we show the column density of synthetic branches compared with Hi-GAL filaments.

\begin{table}
\centering
\begin{tabular}{l c r}
\hline
\hline
Category & Total & Filaments with/no clumps \\
\hline
\textit{Synthetic filaments} & 8,832 & 3,368 (38.1\%) \\
  &  & 5,464 (61.9\%) \\
\hline
\textit{Hi-GAL filaments} & 32,059 & 21,694 (67.7\%) \\
  &  & 10,365 (32.3\%) \\
\hline
\end{tabular}
\caption{Summary of filament statistics: comparison between our synthetic simulation and Hi-GAL observations. We list the total number of filaments, and the number of them with and without embedded clumps. For Hi-GAL, filament counts are from \citet{Schisano+2014ApJ, Schisano+2020MN}. Percentages in parentheses in the last column indicate, for each row, the fraction of filaments that contain clumps (upper value) and those that do not (lower value), both relative to the total number of filaments in that category.}
\label{table_filament}
\end{table}

\subsection{The filaments with embedded clumps}
\label{clump_on_filaments}

In the following, we analyze the spatial association between filaments and clumps to assess the influence of filamentary structures on star formation. In the synthetic catalogue of \citet{Lu-Zu-Jia+2022MNRAS}, compact sources (clumps) were classified as different clump categories following the same method as in the Hi-GAL survey \citep{Elia+2017, Elia+2021MNRAS}. Clumps are distinguished as protostellar or starless clumps based on the presence or absence of a 70~\micron~counterpart, and starless clumps are further separated into gravitationally bound prestellar and unbound categories. The synthetic clump catalogue contains a total of 51,831 clumps with 3D information in \citep{Lu-Zu-Jia+2022MNRAS}. 

\begin{table}
\centering
\begin{tabular}{l c r}
\hline
\hline
Category & Total & Clump on/off filament \\
\hline
\textit{All 2D clumps} & 51,831 & 48,515 (93.6\%) \\
  &  &  3,316 (6.4\%) \\
\hline
Clumps with star & 7,741 (15\%) & 7,285 (94.1\%) \\
embedded in 2D &    & 456 (5.9\%) \\
\hline
Clumps with star & 1,428 (3\%)  & 1,345 (94.2\%) \\
embedded in 3D &   & 83 (5.8\%) \\
\hline
\hline
\textit{All Hi-GAL clumps} & 150,223 & 95,502 (63.6\%) \\
  &  &  54,721 (36.4\%) \\
\hline
\end{tabular}
\caption{Summary of clump statistics: comparison between our synthetic simulation and Hi-GAL observations. For our simulation, we list all 2D synthetic clumps, clumps with an embedded star along the line of sight within the 2D clump radius, and clumps with a star physically embedded in the corresponding 3D clump. The Hi-GAL clump data are from \citet{Schisano+2014ApJ, Schisano+2020MN}. Percentages in parentheses indicate the fraction relative to the total number of clumps in each category, except for the last column, where percentages indicate the fraction within filaments (on) versus outside (off).}
\label{table_clump}
\end{table}

For all the 51,831 synthetic clumps, we find that $94\%$ (48,515 of 51,831) of the clumps are located within the filaments, while only $6\%$ (3,316 of 51,831) lie outside the filaments (see the criteria in Section~\ref{criteria_clump_on_filament}). The histograms of the filament column density in Figure~\ref{fig_6} (blue lines) show that the filaments with embedded clumps peak at $2\times 10^{21}\,\rm{cm}^{-2}$, whereas those without clumps peak at $5\times10^{20}\,\rm{cm}^{-2}$. We find that only $\sim38\%$ (3,368 of 8,832) of filaments have embedded clumps, while $\sim62\%$ (5,464 of 8,832) have no detected clump, suggesting that these filaments may be in an early evolutionary phase.

For all the 150,223 Hi-GAL clumps, there are $64\%$ (95,502 of 150,223) of the clumps are located within the filaments, and $36\%$ of the clumps (54,721 of 150,223) lie outside the filaments (see their criteria in \citet{Schisano+2014ApJ,Schisano+2020MN}). The histograms of the Hi-GAL filament column density in Figure~\ref{fig_6} (black lines) show that the filaments with embedded clumps peak at $5\times 10^{20}\,\rm{cm}^{-2}$, whereas those without clumps peak at $1\times10^{20}\,\rm{cm}^{-2}$. We find that only $\sim68\%$ (21,694 of 32,059) of filaments have embedded clumps, while $\sim32\%$ (10,365 of 32,059) have no detected clump.

Our simulation yields lower background column densities than Hi-GAL (due to the 250 pc box size; see Figure~\ref{fig_5}) but higher filament column densities and a higher clump-embedding fraction: $94\%$ of synthetic clumps are embedded in only $38\%$ of filaments, whereas $64\%$ of Hi-GAL clumps are embedded in $68\%$ of filaments. This result may be attributable to line-of-sight confusion in the observations, which can both elevate the measured background and dilute the contrast of true filamentary structures. Table~\ref{table_filament} summarizes, for both the synthetic and Hi-GAL filament catalogues, the total number of filaments, the counts of those with and without embedded clumps, and the corresponding fractions relative to the total.

\subsection{Clumps with embedded stars}
\label{massive_clump_with_star}

\begin{figure}
\centering  
\includegraphics[width=\linewidth]{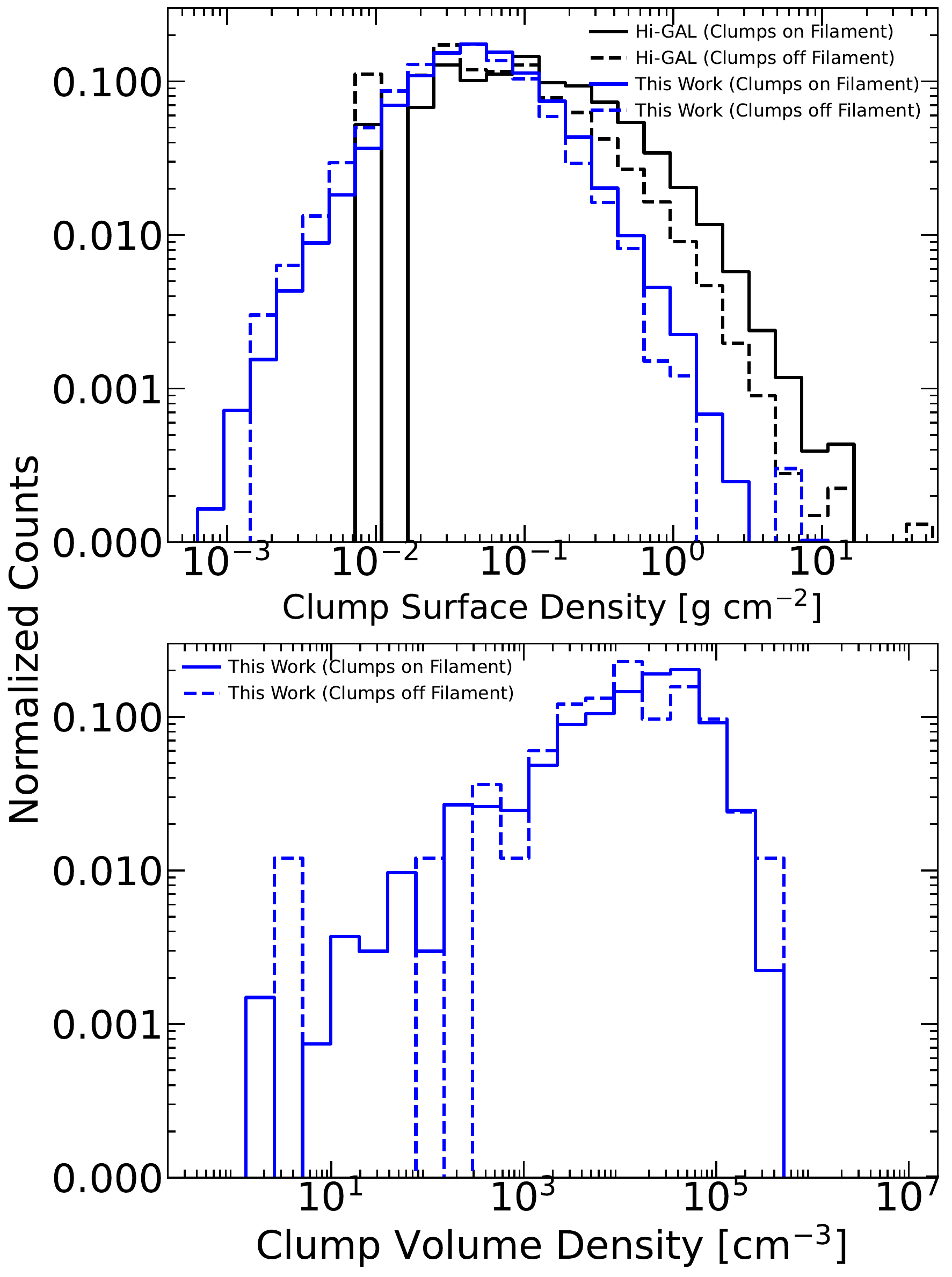} 
\caption{The normalized distributions of all 2D synthetic clump surface densities (top panel) and 3D volume densities of a sub-sample of clumps which have embedded stars in the corresponding 3D clumps from the 3D simulation (bottom panel). The blue lines represent the clumps from our simulation, while the black lines (top panel only) represent the clumps from the Hi-GAL survey. Solid lines denote clumps located on filaments, while dashed lines denote clumps located off filaments.
}
\label{fig_7}  
\end{figure}

For all 51,831 2D synthetic clumps, we match them with star positions from the 3D MHD simulation to determine whether a star is located within the 2D clump radius along the line of sight. We find that 7,741 of the 51,831 ($\sim 15\%$ of all) clumps host an embedded star, regardless of whether the star lies inside the 3D clump or outside it but along the line of sight of the 2D synthetic clump. Only 1,428 of the 51,831 ($\sim 3\%$ of all) 2D clumps have detected a star in the corresponding 3D clump.

For the 7,741 2D synthetic clumps that have an embedded star along the line of sight, $\sim94\%$ (7,285 of 7,741) lie on filaments, while $\sim6\%$ (456 of 7,741) are located in off-filament regions.  For the 1,428 2D synthetic clumps that have an embedded star in the corresponding 3D clump, the ratios of the clumps on and off the filaments are also $\sim94\%$ (1,345 of 1,428) and $\sim6\%$ (83 of 1,428), respectively. The two sub-samples of synthetic clumps - namely, clumps that have an embedded star along the line of sight within the 2D clump radius, and clumps that have an embedded star in the corresponding 3D clump - exhibit the same on/off filament ratio as the whole clump sample. We summarize the full synthetic clumps and the two sub-samples in Table~\ref{table_clump}.

To evaluate the physical conditions of the clumps in different environments, we examine the surface densities ($\Sigma$) of the clumps. $\Sigma$ is derived with $\Sigma = M / (\pi R^2)$, where $M$ and $R$ are the 2D mass and radius of the synthetic clump.

In the top panel of Figure~\ref{fig_7}, we show the clump surface density of all the synthetic (blue) and Hi-GAL (black) clumps. The synthetic clumps within the filaments exhibit higher typical $\Sigma$ values, with a median of $0.07\,\rm{g\,cm}^{-2}$ and reaching up to $4.28\,\rm{g\,cm}^{-2}$, compared to the clumps outside of the filaments, which have a lower median of $0.05\,\rm{g\,cm}^{-2}$ and do not exceed $0.6\,\rm{g\,cm}^{-2}$ (a two-sample K-S test on the $\Sigma$ distributions of the on-filament and off-filament clump populations yield $D = 0.20$, $p<0.001$). This trend is consistent with the Hi-GAL data (black lines), where filament-associated clumps also exhibit higher median surface densities ($0.11\,\rm{g\,cm^{-2}}$) compared to off-filament ones ($0.06\,\rm{g\,cm^{-2}}$). However, the contrast is less pronounced in our simulation compared to the Hi-GAL results. Most of the clumps from both the synthetic and real observations lie below the surface density threshold of $1\,\rm{g\,cm}^{-2}$ suggested for massive star formation by \citet{Krumholz+2008Nature}, or the revised threshold of $\boldsymbol{\sim} 0.2\,\rm{g\,cm}^{-2}$ by \citet{Butler&Tan+2012ApJ}.

In the bottom panel of Figure~\ref{fig_7}, we show the 3D volume density of a sub-sample clumps, which has an embedded star in the corresponding 3D clumps. The volume density is calculated within the size that assuming the line-of-sight depth equal to the projected diameter of each clump with the 3D simulation data as presented in \citep{Lu-Zu-Jia+2022MNRAS}. The clumps located within the filaments exhibit a higher typical density, with a median of $2.3\times10^4\,\rm{cm}^{-3}$, which is approximately $55\%$ higher than the median density of $1.5\times10^4\,\rm{cm}^{-3}$ for isolated off-filament clumps (K-S test: $D=0.20$, $p<0.001$). Furthermore, $94\%$ (919 out of 973) of clumps that exceed the star formation threshold of $10^4\, \mathrm{{cm}^{-3}}$ \citep{Lada-Elizabeth+1993prpl} are associated with filaments, underscoring the importance of filaments as dense structures that form stars.

\begin{figure} 
\centering  
\includegraphics[width=\linewidth]{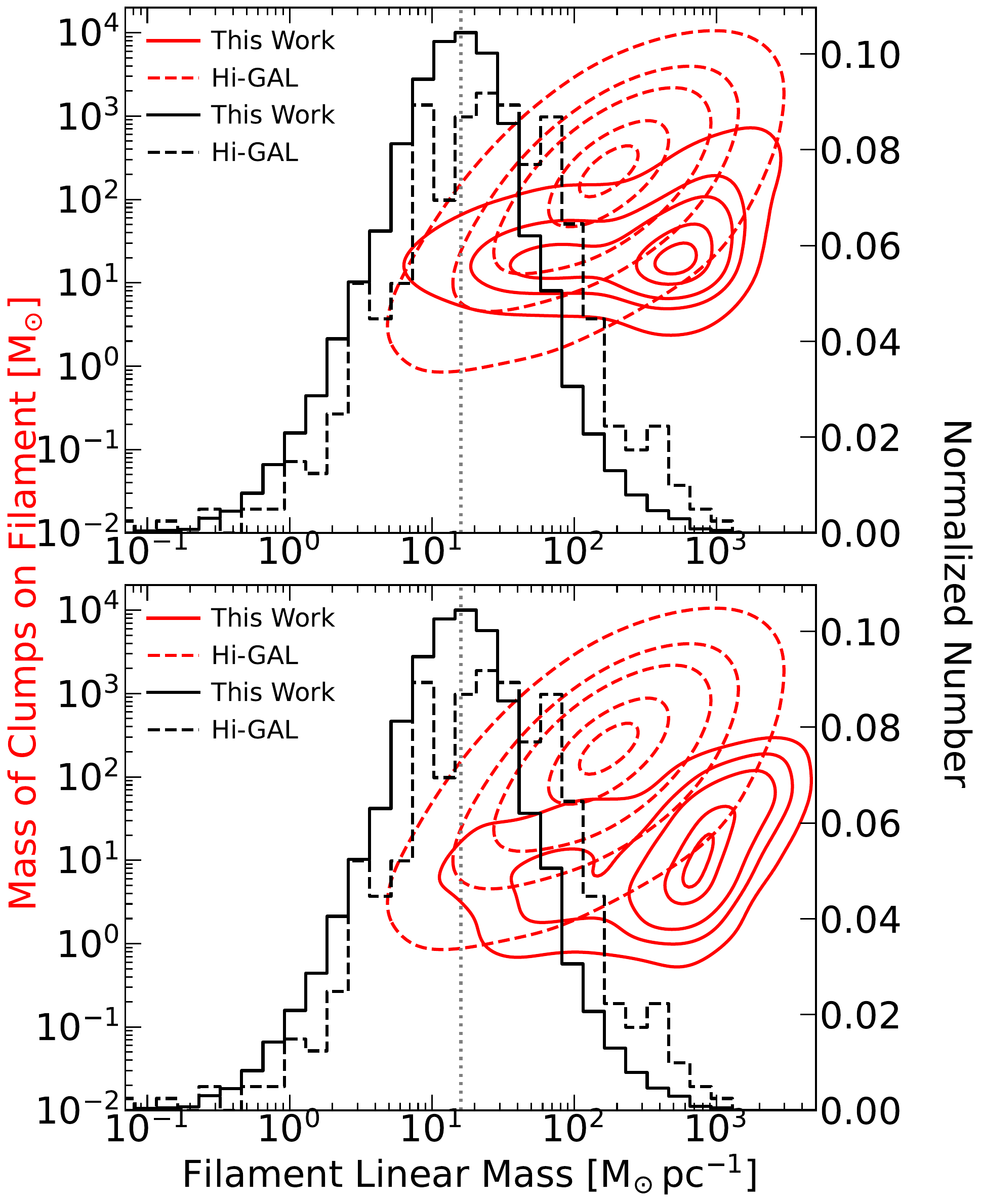}  
\caption{\textit{Top panel:} relationship between clump mass (left y-axis) and the line mass ($M_{\rm line}$) of their host structures. The red solid contours represent synthetic filaments, while the red dashed contours show the filaments from the Hi-GAL survey \citep{Schisano+2020MN}. The contour levels correspond to $90\%$, $70\%$, $50\%$, $20\%$, and $5\%$. The vertical dotted line marks the critical line mass ($16\,\rm M_\odot\,pc^{-1}$ at $T=10\,\rm K$). The black solid histogram (right y-axis) shows the distribution of $M_{\rm line}$ for synthetic filaments without clumps, compared to the black dashed line representing the Hi-GAL filaments without clumps. \textit{Bottom panel:} the same as the top panel, but for clumps with embedded stars (mass derived from 3D gas).}
\label{fig_8}  
\end{figure}

\subsection{The mass of filaments}\label{mass_of_filaments}

The stability of filaments depends on their linear density or mass per unit length ($M_{\rm{line}}$), which also governs their capacity to form stars. In a cylindrical model, $M_{\rm{line}}>M_{\rm{line, crit}}$ initiates radial collapse under self-gravity, which produces dense cores \citep{Ostriker+1964ApJ, Larson+1985MNRAS, Andre.P+2010A&A}. For $T\sim10\,\rm{K}$, $M_{\rm{line,crit}}\approx16\,\rm{M}_\odot\,\rm{pc}^{-1}$, modestly elevated by turbulence or magnetic fields \citep{Fiege+2000MNRAS}.

\begin{figure}
\centering  
\includegraphics[width=\linewidth]{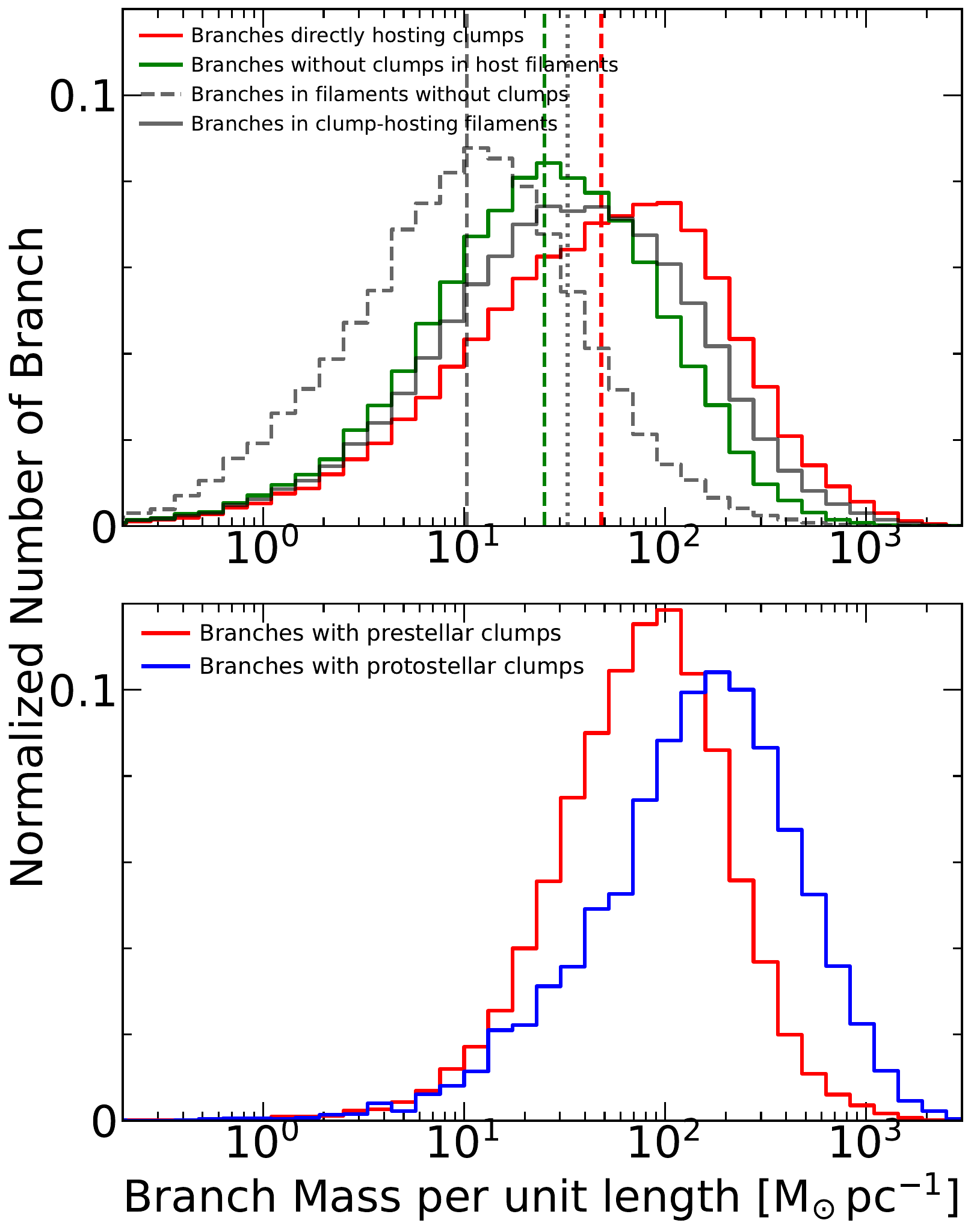}  
\caption{\textit{Top panel:} distributions of the line mass ($M_{\rm line}$) for individual branches. The black solid and dashed lines are the branches corresponding to the filaments with and without clump embedded. Branches belonging to the clump-hosting filaments are further separated into two subsets: those directly hosting clumps (red) in branches and those without clumps (green) in branches but have clumps embedded in their connected filaments. Vertical lines indicate the median value of each distribution. \textit{Bottom panel:} distributions of $M_{\rm line}$ for the specific subset of branches hosting prestellar clumps (red) and protostellar clumps (blue).}
\label{fig_9}  
\end{figure}

Based on the column density, the masses of the filaments and their constituent branches were calculated using the method proposed by \citet{Ma+2023A&A},
\begin{equation}
M=\mu \, m_{\mathrm{H}} \, \sum_iA_{\mathrm{pixel}}(i) \, N_{\mathrm{H_2}}(i)\,,
\end{equation}
where \(A_{\mathrm{pixel}}(i)\) represents the area of a pixel and \(N_{\mathrm{H_2}}(i)\) is the hydrogen molecule column density corresponding to the pixel.

The top panel of Figure~\ref{fig_8} illustrates the masses of clumps identified within the filaments as a function of $M_{\rm{line}}$, comparing our filaments (solid contours) with the filament sample from the Hi-GAL survey (dashed contours; \citealt{Schisano+2020MN}). We exclude filaments containing starless clumps from our analysis. Both datasets exhibit a positive correlation, which is physically expected, as filaments with higher line masses provide deeper gravitational potential wells for fragmentation into clumps \citep{Andre.P+2010A&A, Andre+2014prpl, Arzoumanian+2011A&A}. In Appendix \ref{app:branch_properties}, we show the same relations as Figure~\ref{fig_8} in Figure~\ref{fig_app_linear_mass}, but for the synthetic branches compared with the Hi-GAL filaments.

Building upon the mass correlation shown in Figure~\ref{fig_8}, we further examine the $M_{\rm line}$ distribution across the constituent branches. In Figure~\ref{fig_9}, we categorize these branches based on the presence of clumps. The top panel presents the distributions of $M_{\rm line}$ for filaments belonging to filamentary networks with and without clumps. Filaments in clump-hosting filamentary networks exhibit significantly higher line masses (median $48.1\,\rm{M}_\odot\,\mathrm{{pc}^{-1}}$) than those in clump-free networks (median $10.3\,\mathrm{{M}_\odot\,\mathrm{{pc}^{-1}}}$). Even individual filaments without clumps, but located within clump-hosting networks, tend to have elevated $M_{\rm line}$ values (median $25.0\,\rm{M}_\odot\,\mathrm{{pc}^{-1}}$), suggesting that supercritical conditions can extend beyond immediate star-forming sites. The bottom panel separates the clump-hosting branches into prestellar and protostellar types. The protostellar filaments ($19\%$) exhibit a higher median $M_{\rm line}$ of $154.0\,\rm{M}_\odot\,\mathrm{{pc}^{-1}}$ compared to $78.0\,\rm{M}_\odot\,\mathrm{{pc}^{-1}}$ for the prestellar filaments ($81\%$). This trend is consistent with an evolutionary sequence in which an increasing line mass facilitates the transition from the prestellar to the protostellar stage.

\subsection{Stars and filamentary structures}
\label{Stars_and_Filamentary_Structures}

\begin{figure}
\centering  
\includegraphics[width=\linewidth]{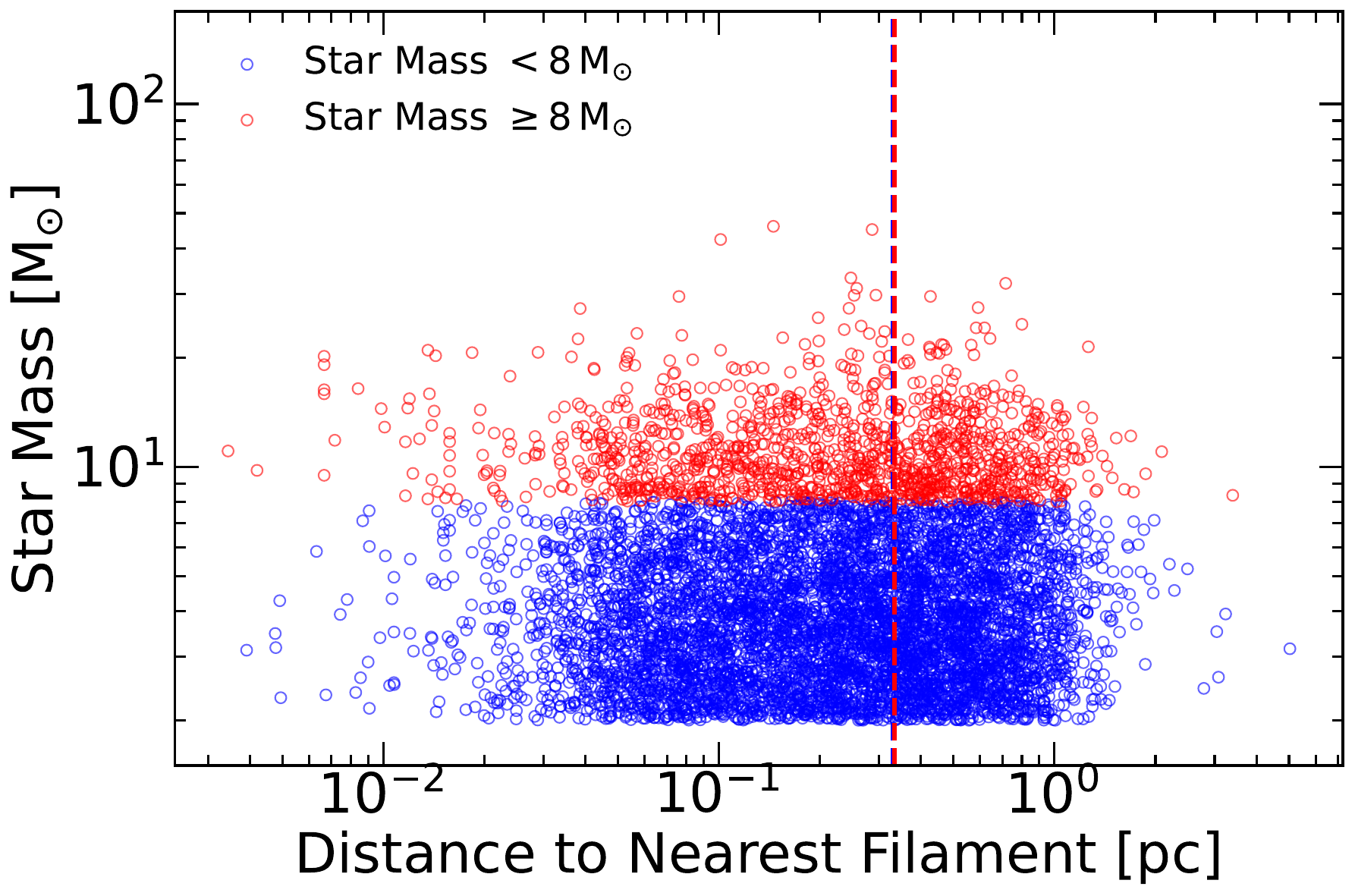}  
\caption{Stellar mass versus distance to filaments. The red dots are the mass of massive stars. The vertical dashed lines mark the average distances to the filaments, which are nearly identical at 0.33~pc for both massive (red) and low-mass (blue) stars.}
\label{fig_10}  
\end{figure}

By combining filament and clump analyses, we suggest that filaments can guide matter into dense clumps and promote star formation. To further quantify the role of filaments in massive star formation, we analyze the spatial association between stars and filaments.

Filaments critically shape both the spatial distribution of stellar populations and the accretion process that builds up the stellar mass, particularly for massive stars. Overall, 93\% of stars (6,681 of 7,208) are located within filaments, reinforcing the role of filaments in facilitating gravitational collapse and material accretion. Among these, 92\% of massive stars (1,178 of 1,287) are embedded within filaments. If massive stars were distributed randomly across the simulated region, their expected occurrence rate within filaments would correspond to the area fraction of the filamentary networks. By calculating the exact pixel ratio from our 36 \texttt{FILFINDER} extraction masks, we determine this expected random occurrence rate to be $38.5\% \pm 3.4\%$,  which is significantly lower than the observed 92\% occurrence rate for massive stars. Figure~\ref{fig_10} shows that stellar mass decreases with increasing distance from the filament. For $75\%$ of the massive stars, the projected distance to the nearest filament is less than 0.47~pc. Interestingly, both massive and lower-mass stars are tightly clustered around the filaments, sharing a similarly small average projected distance of 0.33~pc.

\section{Discussion} \label{Discussion}

\subsection{Connecting 2D synthetic observables to underlying 3D physics}
\label{Connecting_2D_Synthetic_Observables_with_Underlying_3D_Physics}

Connecting 2D synthetic observables with the underlying 3D physics allows us to assess how well intrinsic physical relationships are preserved or obscured in observational data. However, interpreting 2D projected column density maps remains challenging due to projection effects and line-of-sight confusion, which can generate spurious features or bias derived properties \citep{Panopoulou+2017MNRAS}. To bridge 2D observables and 3D physics, we compare our synthetic filament properties with previous 3D numerical simulations. 

The environmental scaling relationship ($N_{\rm fs}\propto N_{\rm bs}^{0.78}$) identified in our synthetic maps (see Section~\ref{column_density_properties_of_filaments}) reflects filament formation driven by the shock-compression of dense turbulent flows \citep{Federrath+2016MNRAS, Seifried+2020MNRAS}. 
In these models, filaments represent the dense post-shock gas generated by converging supersonic flows. Under typical isothermal shock conditions, this post-shock density is correlated with the pre-shock ambient density \citep{Padoan+1997MNRAS}. Consequently, shock compression in denser background environments systematically produces denser filaments, establishing the observed positive correlation. The flatter slope observed here relative to observational surveys suggests that line-of-sight confusion in Galactic-scale studies likely steepens this correlation. Furthermore, 2D-identified supercritical filaments correlate with 3D gravitational instability \citep{Clarke+2017MNRAS}. Their spatial alignment with collapse events \citep{Zucker+2018ApJ}, alongside the extreme concentration of clumps ($94\%$) and massive stars ($92\%$) within them, confirms that filaments function as the primary channels for mass accretion and hierarchical fragmentation \citep{Gomez+2014ApJ, Smith+2016MNRAS, Padoan+2020ApJ}.

\subsection{The physical nature and evolution of filaments}
\label{The_Physical_Nature_and_Evolution_of_Filaments}

\begin{figure}
\centering  
\includegraphics[width=\linewidth]{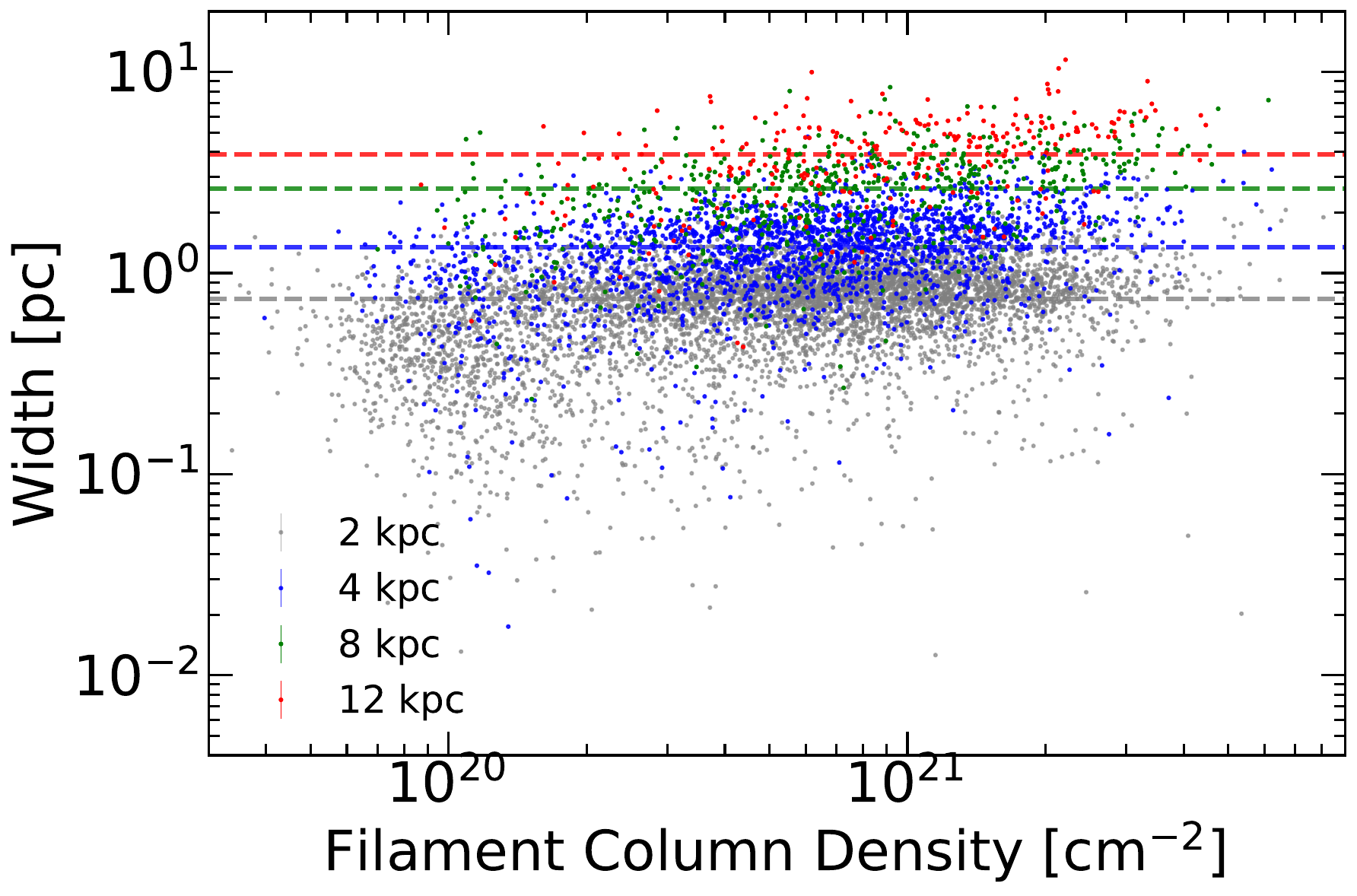}  
\caption{The relation between filament widths and central column densities at different distances. Individual filaments are represented by filled circles. The horizontal dashed lines denote the median widths of the corresponding distance groups: 2~kpc (grey), 4~kpc (blue), 8~kpc (green), and 12~kpc (red).}
\label{fig_11}  
\end{figure}

Our derived statistical properties capture key signatures of environmentally driven filament evolution. Building on the environmental scaling discussed above, this positive correlation empirically indicates that dense filaments preferentially reside in higher-density environments \citep{Schisano+2014ApJ}.

Beyond formation, their dynamical state undergoes a clear transition toward gravitational dominance. As demonstrated in 3D radiative transfer simulations \citep{Seifried+2017MNRAS, Seifried+2020MNRAS}, magnetic fields reorient perpendicular to the filament axis at densities of $n_{\rm{trans}}\sim10^2\text{--}10^3\,\rm{cm}^{-3}$ as the mass-to-flux ratio approaches unity, triggering clump formation. Thus, exceeding the gravitational supercriticality threshold marks the decisive phase where filaments destabilize, fragment, and initiate star formation. 

Despite this gravitational dominance, a statistical analysis of our overall sample yields a weak positive correlation (Spearman rank correlation test: $\rho = 0.31$, $p<0.001$; Figure~\ref{fig_11}), indicating no clear dependence of filament width on column density. This finding is consistent with the nearly constant widths reported by \citet{Arzoumanian+2011A&A} in their study of filament sizes in nearby star-forming regions. However, this result must be interpreted with caution, as the finite pixel resolution in our synthetic maps (e.g., $0.11$\,pc at $d=2$\,kpc, and coarser at larger distances) imposes an artificial lower limit on the measurable width. This resolution limit inherently flattens the true underlying slope by preventing us from resolving narrower structures at high densities. Nevertheless, if the physical widths are indeed stable against radial contraction, this stability likely arises from accretion-driven turbulence \citep{Heigl+2020MNRAS}, where accretion energy continuously converts into internal turbulent pressure \citep{Federrath+2016MNRAS}, maintaining constant widths even at supercritical line masses \citep{Clarke+2017MNRAS}.

\subsection{Filaments as engines of star formation}
\label{Filaments_as_Engines_of_Star_Formation}

\begin{figure}
\centering  
\includegraphics[width=\linewidth]{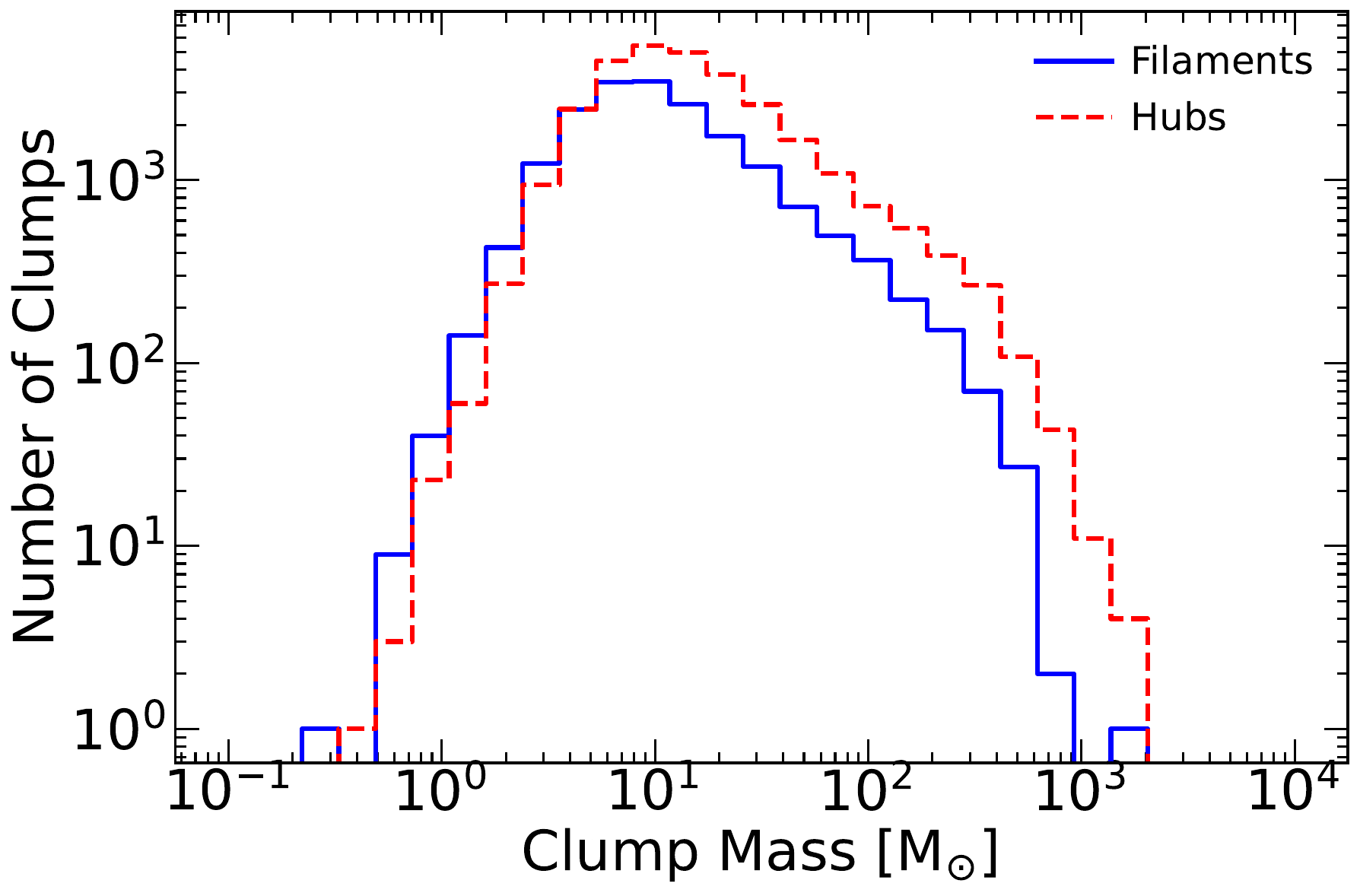}  
\caption{Clump distribution in HFS: blue denotes clumps associated with filaments; red indicates hub-located clumps.}  
\label{fig_12}  
\end{figure}

\begin{figure*}
\centering  
\includegraphics[width=1.0\linewidth]{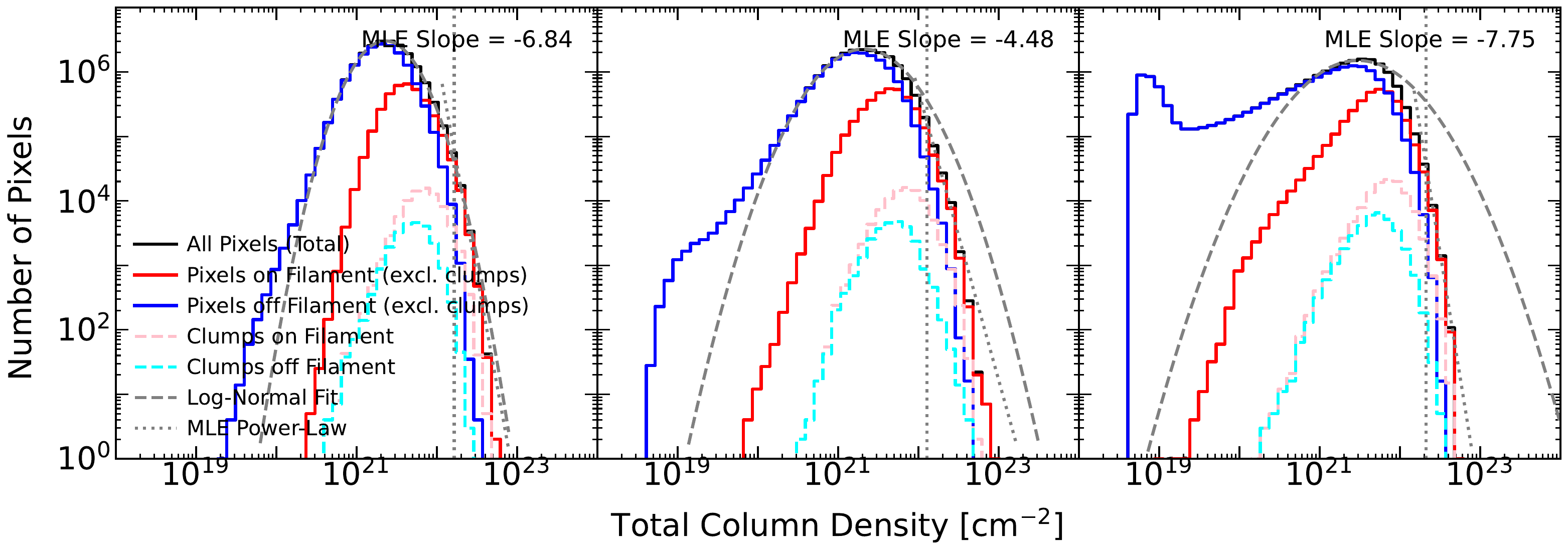}
\caption{Column density PDFs at times 15.4, 23.3, and 34.2~Myr (from left to right) illustrating the density evolution in the simulated molecular clouds. The grey dashed lines represent the log-normal fits. The grey dotted lines denote the MLE power-law fits, yielding slopes of $-6.84$, $-4.48$, and $-7.75$, respectively. The vertical dotted lines mark the lower limits ($x_{\min}$) adopted for the power-law fitting.}
\label{fig_13}
\end{figure*}

The embedding of $94\%$ of clumps and $93\%$ of stars ($92\%$ of massive stars) within filaments provides robust quantitative evidence for the column density threshold model \citep{Andre+2014prpl}. Filaments must exceed a critical surface density ($\Sigma_{\rm th} \sim 130\,\mathrm{M}_{\odot}\,{\rm pc}^{-2}$, as proposed by \citealt{Andre+2014prpl}) to trigger prestellar core fragmentation, thereby serving as the necessary prerequisite environment for high-mass star formation \citep{Molinari+2010A&A}.

Rather than isolated core formation, this tight clump-filament connection highlights filaments as dynamic conduits that continuously transport gas from molecular clouds down to dense clumps \citep[e.g.,][]{Gomez+2014ApJ, Smith+2016MNRAS, Vzquez-Semadeni+2019MNRAS}. Continuously fed by these longitudinal accretion flows, clumps grow massive enough to form high-mass stars before stellar feedback halts the accretion \citep{Chen+2019ApJ, Padoan+2020ApJ}. Conversely, high-density filament segments devoid of clumps represent the prestellar phase, actively accumulating mass via turbulent compression prior to widespread fragmentation
, a behavior consistent with radial accretion flows onto filaments \citep{Kirk+2013ApJ}. Thus, our sample successfully traces the full evolutionary sequence from initial gas assembly to the formation of supercritical massive clusters. Indeed, the most extreme of these star-forming hubs do not evolve within isolated cylindrical structures but are dynamically fueled by the intersection of multiple converging channels.

\subsection{The role of hub-filament systems in star formation}
\label{The_Role_of_Hub-Filament_Systems_in_Star_Formation}

In this work, we focus on the properties of individual filaments. However, the most extreme manifestations of mass accretion occur where multiple structures intersect. Individual channels naturally converge in dense star-forming regions to form hub-filament systems  \citep[e.g.,][]{Myers+2009ApJ, Peretto+2013A&A, Kumar+2020A&A}. These HFSs serve as fundamental sites for massive star formation, where converging filaments create deep gravitational wells for the rapid accumulation of gas \citep[e.g.,][]{Myers+2009ApJ, Peretto+2013A&A, Motte+2018ARA&A, Trevio-Morales2019A&A, Kumar+2022A&A}.

Our data strongly reinforce filaments as primary mass delivery mechanisms. Notably, more than 50,000 identified clumps reside within filaments, and over $60\%$ of these are concentrated directly within the central hub, as shown in Figure~\ref{fig_12}. The higher average mass of the clumps located in the hub ($30.9\,\rm{M}_{\odot}$, with a maximum mass of $1,719\,\rm{M}_{\odot}$) compared to those distributed along the surrounding filaments ($21.2\,\rm{M}_{\odot}$, maximum $1,819\,\rm{M}_{\odot}$) suggests an enhanced efficiency of the hub in gathering dense gas. This mass discrepancy is statistically significant (K–S test: $D = 0.144$, $p < 0.001$) and is consistent with observational findings of HFS density amplification \citep{Kumar+2020A&A}. Furthermore, the fraction of the total clump mass across the entire simulation map that is concentrated within the hubs increases steadily between 15.4 and 34.2~Myr. This evolution toward a centrally concentrated mass distribution is consistent with structural trends in ALMAGAL clumps \citep{Schisano+2026apj}, indicating that hubs actively accumulate mass via filamentary transport driven by global collapse \citep{Peretto+2013A&A}. These findings align with recent observations of parsec-scale cluster-forming hubs, where high gas surface densities ($\Sigma_{\rm gas}$) yield extreme free-fall efficiencies ($\epsilon_{\rm ff}\sim 13\%$) well above standard Kennicutt-Schmidt relations \citep{Rawat+2025MN}.

\subsection{Limitations and implications for synthetic observations}
\label{Limitations_and_Implications_for_synthetic_Observationals}

Our synthetic column density PDFs deviate from the log-normal distributions with prominent power-law tails typically observed in evolved turbulent clouds \citep[e.g.,][]{Kainulainen+2009A&A,Kainulainen+Tan13,Schneider+2013ApJ,Schneider+2022A&A}. Although the density distribution peak shifts higher over time as shown in Figure~\ref{fig_13}, no distinct power-law tail emerges.

This absence stems from both the cloud evolutionary phase and projection effects. Physically, supernova-driven turbulence in the simulation redistributes gas via uniform compression rather than the localized hierarchical collapse responsible for complex high-density tails \citep{Andre+2014prpl, Pillsworth+2024MNRAS}. 
Observationally, 2D projections smooth line-of-sight density contrasts, obscuring localized collapse signatures that are otherwise easily resolved in 3D volumes or advanced stages \citep[e.g,][]{Smith+2016MNRAS, Zamora+2017MNRAS, Auddy+2019ApJ, Seifried+2020MNRAS}. This suggests our simulated filaments remain in a turbulence-dominated phase where global gravity has not yet fully superseded turbulent compression, consistent with multiscale assembly models \citep{Padoan+2020ApJ}.

The choice of structure identification algorithm introduces inherent morphological biases that affect 2D geometric comparisons. For example, our flatter length distribution relative to the Hi-GAL catalog ($\alpha_L\approx-1.7$ vs. $-4.0$) directly reflects this algorithmic dependence. The \texttt{FILFINDER} approach connects adjacent structures into continuous backbones via skeletonization \citep{Koch+2015MNRAS}, whereas the Hessian-based Hi-GAL method \citep{Schisano+2014ApJ, Schisano+2020MN} fragments networks into shorter segments at localized curvature variations. Although algorithmic differences inevitably introduce morphological variations, integrated quantities such as the filament line mass remain reliable diagnostics of the underlying gravitational state.

\section{Conclusions} 
\label{Conclusion}

In this study, we identify the filaments from the column density maps, which are computed from our synthetic Herschel observations \citep{Lu-Zu-Jia+2022MNRAS} derived from large-scale 3D MHD simulations \citep{Padoan+SN1+2016ApJ}. It results in 8,832 filaments, which are decomposed into 110,193 branches. We analyze the physical properties of the filaments and clumps from the simulations, and compare our filaments with the catalogue of 32,059 Galactic filaments from the Hi-GAL survey presented in \citet{Schisano+2020MN}. Here we summarize the main findings of the work in the following.

(i) We find power-law distributions for our synthetic filament masses and lengths, with power-law index of $\alpha_{\rm M}=-0.86$ and $\alpha_{\rm L} = -1.71$, respectively. It is similar to Hi-GAL observations in \citet{Schisano+2020MN} and the values of our power-law index are consistent with the filaments extracted directly from column density of galactic-scale MHD simulations in \citet{Pillsworth+2025ApJ}, which used the same filament extraction algorithm \texttt{FILFINDER}.

(ii) We find that filaments act as the primary reservoirs for star formation, hosting $94\%$ of clumps and $93\%$ of stars ($92\%$ of massive stars, $M \gtrsim 8\,\rm M_{\odot}$). Notably, filaments hosting clumps exhibit significantly higher median column densities ($1.1\times10^{21}\,\rm{cm}^{-2}$) compared to those without ($3.8\times10^{20}\,\rm{cm}^{-2}$), a result consistent with the existence of a column density threshold for efficient fragmentation.

(iii) We also find a distinct environmental scaling relation, $N_{\rm{fs}} \propto N_{\rm{bs}}^{0.78}$, which supports a scenario that filament formation is driven by the shock compression of turbulent flows. The majority of these structures are gravitationally supercritical, which have $M_{\rm line}$ above the critical value of 16 $\rm M_{\odot}\,\rm{pc}^{-1}$. These findings are consistent with models where filaments function as accretion channels feeding dense hubs. The flatter slope of our scaling relation compared to Galactic surveys implies that line-of-sight confusion in real observations likely artificially steepens this intrinsic correlation.

We assess the reliability of 2D projected metrics in tracing the intrinsic 3D physical state of the filaments. Although projection effects and the early evolutionary stage of the simulation suppress certain features, such as the power-law tails in column density PDFs, key statistical signatures, including density scaling and the spatial association between filaments and clumps, remain robust. This demonstrates that, despite observational limitations, 2D synthetic observations provide an effective baseline for interpreting the underlying dynamics of mass accretion and fragmentation in large-scale surveys.

\section{Acknowledgments} \label{Acknowledgments}

We thank the anonymous referee for their constructive review, which significantly improved the quality of the paper. We thank Paolo Padoan and Veli-Matti Pelkonen for providing comments on the manuscript. We also thank Eugenio Schisano for sharing the observation data. This research is supported by Guangxi Natural Science Foundation under Grant No.AD23026127, 2024GXNSFBA010436. ZJL acknowledges support by NSFC grant No.12563005, No.12494571 and the National Key Research and Development (R$\&$D) Program of China No.2024YFA1611704. This research is also supported by Guangxi Qingmiao Talent Support Program, Guangxi Key Research and Development Program (Guike FN2504240040), Bagui Scholars Programme (W.X.-G., GXR-6BG2424001), Guangxi Talent Program (``Highland of Innovation Talents”).

\restartappendixnumbering
\appendix
\onecolumngrid

\section{Comparison of synthetic branches with filaments from Hi-GAL Survey}
\label{app:branch_properties}

\begin{figure*}[h!]
\centering  
\includegraphics[width=1.0\linewidth]{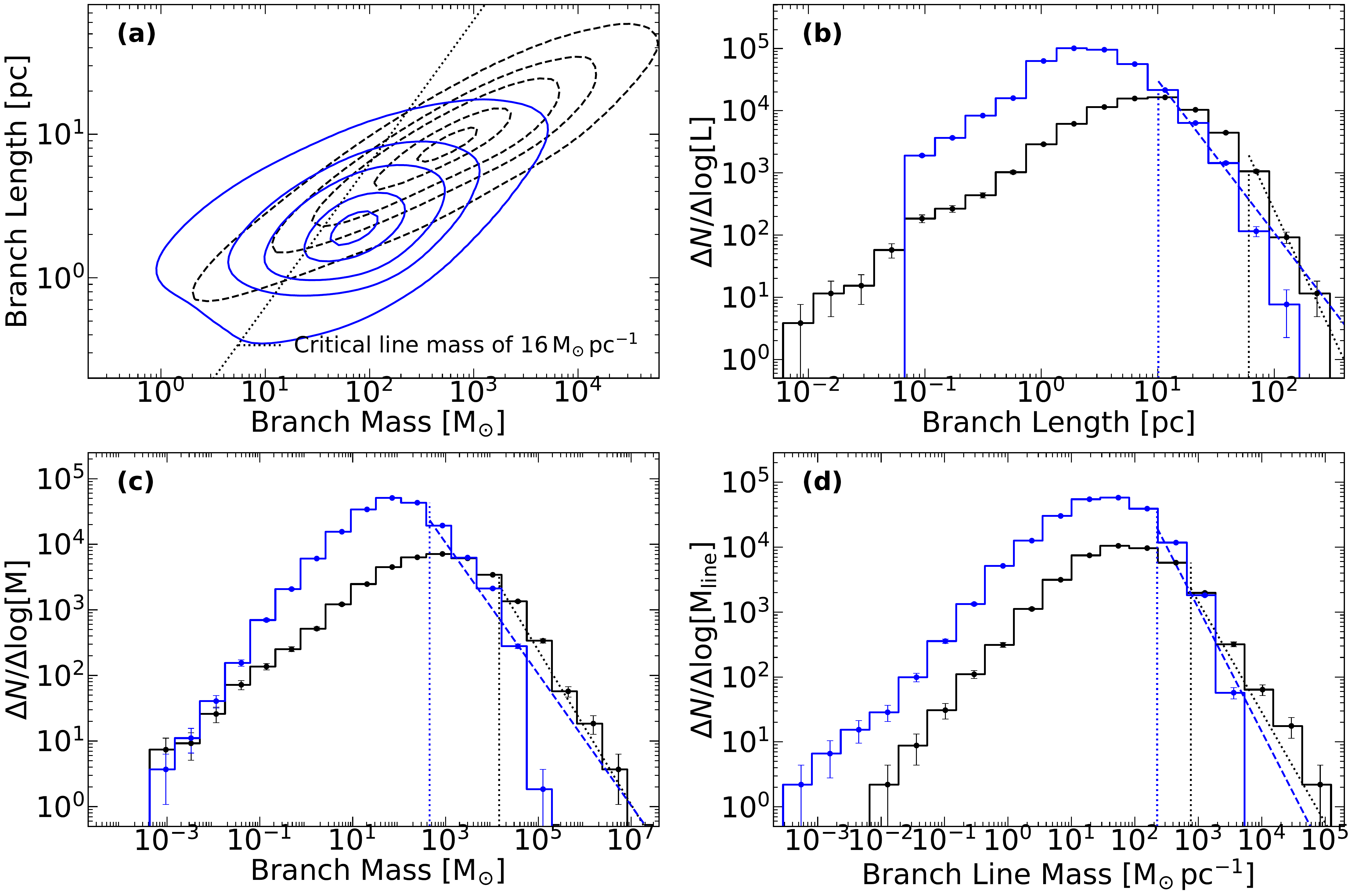}
\caption{Properties of synthetic branches vs.\ Hi-GAL filaments \citep{Schisano+2020MN}. \textit{(a)} Length vs.\ mass. Contours (blue for synthetic branches, black for Hi-GAL) indicate $90\%$, $70\%$, $50\%$, $20\%$, and $5\%$ of the maximum number density. The dashed line is the critical line mass of $16\,\rm M_\odot\,pc^{-1}$ at $T = 10\rm\,K$. \textit{(b)-(d)} Distributions of length, background-subtracted mass, and line mass. Solid histograms (blue for synthetic branches, black for Hi-GAL) include Poisson errors ($\sqrt{N}$). Blue dashed (synthetic) and black dotted (Hi-GAL) lines show MLE power-law fits above the lower limits ($x_{\min}$, vertical dotted lines). Fitted slopes (This work vs.\ Hi-GAL): (b) $\alpha_{\mathrm{L}} = -2.45 \pm 0.03$ vs.\ $-4.02 \pm 0.34$; (c) $\alpha_{\mathrm{M}} = -1.00 \pm 0.01$ vs.\ $-1.17 \pm 0.04$; (d) $\alpha_{\mathrm{M}_{\rm line}} = -1.90 \pm 0.02$ vs.\ $-1.70 \pm 0.06$.}
\label{fig_app_four_panel}
\end{figure*}

In the appendix, we show the comparisons with observations but for our synthetic branches. Figure~\ref{fig_app_four_panel}, the same as Figure~\ref{fig_3}, shows the physical properties of synthetic branches compared with the filaments from the Hi-GAL survey \citep{Schisano+2020MN}. Panel (a) of Figure~\ref{fig_app_four_panel} shows that the majority of our massive branches ($\gtrsim 100\,\mathrm{M}_\odot$) are gravitationally supercritical, lying below the thermal critical line mass of $16\rm \,M_\odot\,\rm{pc}^{-1}$. The distributions of length, mass, and line mass (panels b, c, d) exhibit power-law tails ($\mathrm{d}N/\mathrm{d}\log{X}\propto X^{\alpha}$) fitted using the Maximum Likelihood Estimation method \citep{Bilous+2025A&A}. While our simulated branches show a shallower length distribution than the Hi-GAL sample ($\alpha_{\mathrm{L}} = -2.45 \pm 0.03$ vs.\ $-4.02 \pm 0.34$), their line mass distributions are comparable ($\alpha_{M_{\rm line}} = -1.90 \pm 0.02$ vs.\ $-1.70 \pm 0.06$). The line masses of our synthetic branches span an extensive range from $\sim 10^{-3}$ to $10^4\,\rm{M}_\odot\,\rm{pc}^{-1}$. The high-mass tail of this distribution (${\gtrsim}200\,\rm{M}_\odot\,\rm{pc}^{-1}$) is consistent with the massive filamentary structures observed in the ATLASGAL survey \citep{Li-Guang-Xin+2016A&A} and the nearby molecular clouds \citep{Andre+2019A&A}.

\begin{figure}
\centering  
\includegraphics[width=\columnwidth]{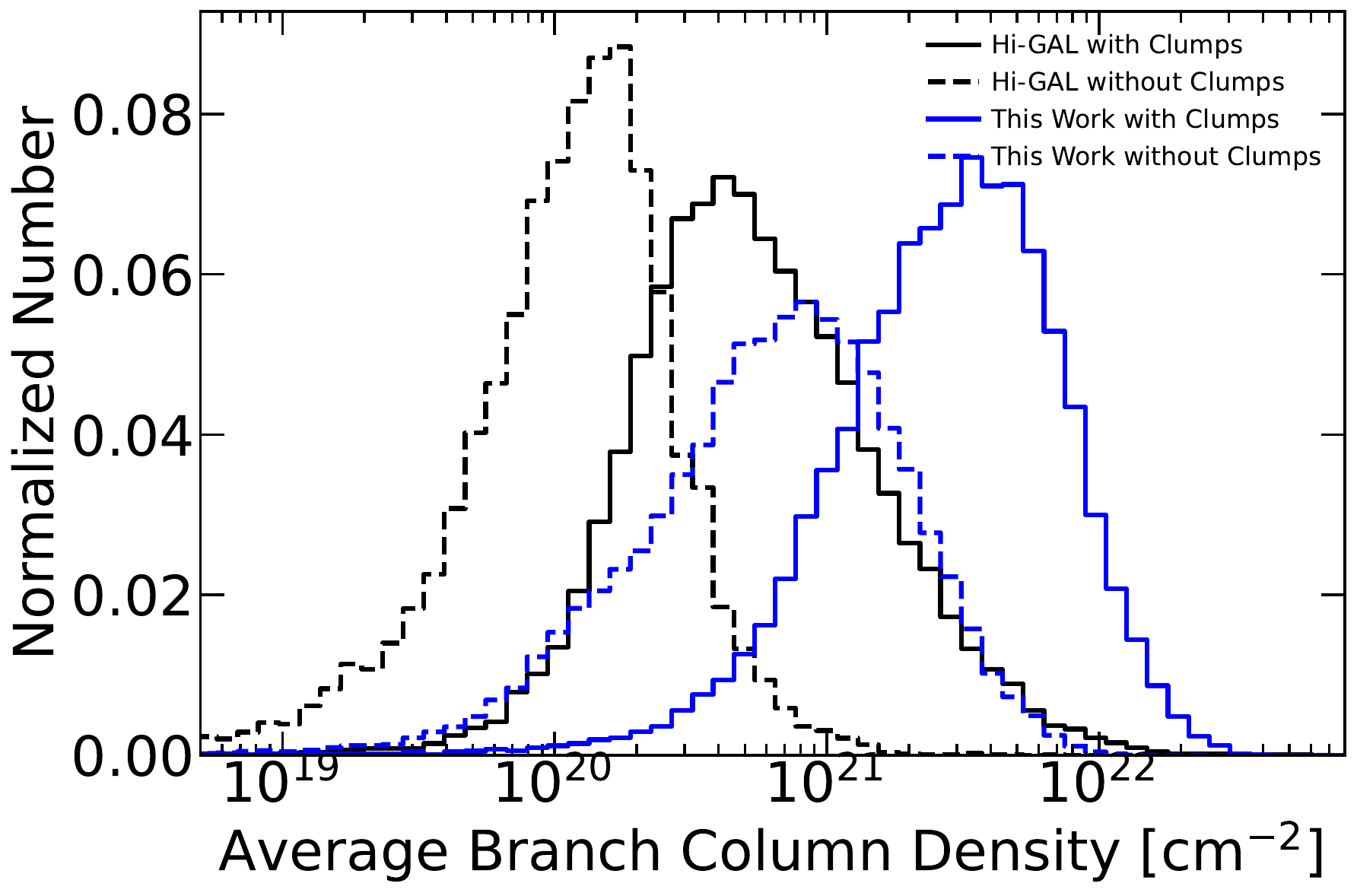}  
\caption{Column density histograms for synthetic branches (blue lines) and the Hi-GAL filaments from \citet{Schisano+2020MN} (black lines). Solid lines indicate structures associated with clumps, while dashed lines indicate those without clumps.}  
\label{fig_app_col_dens}  
\end{figure}

In Figure~\ref{fig_app_col_dens}, we compare the column densities of synthetic branches with Hi-GAL filaments, which are categories that are embedded with and without clumps. While the Hi-GAL samples show a distinct separation between the two populations, our synthetic branches exhibit larger overlap due to the globally elevated density of the simulated hub environment. Nevertheless, our branches associated with clumps still maintain a statistically higher median column density ($2.8\times10^{21}\,\rm{cm}^{-2}$) than those without clumps ($5.2\times10^{20}\,\rm{cm}^{-2}$). 

\begin{figure}[t]
\centering  
\includegraphics[width=\columnwidth]{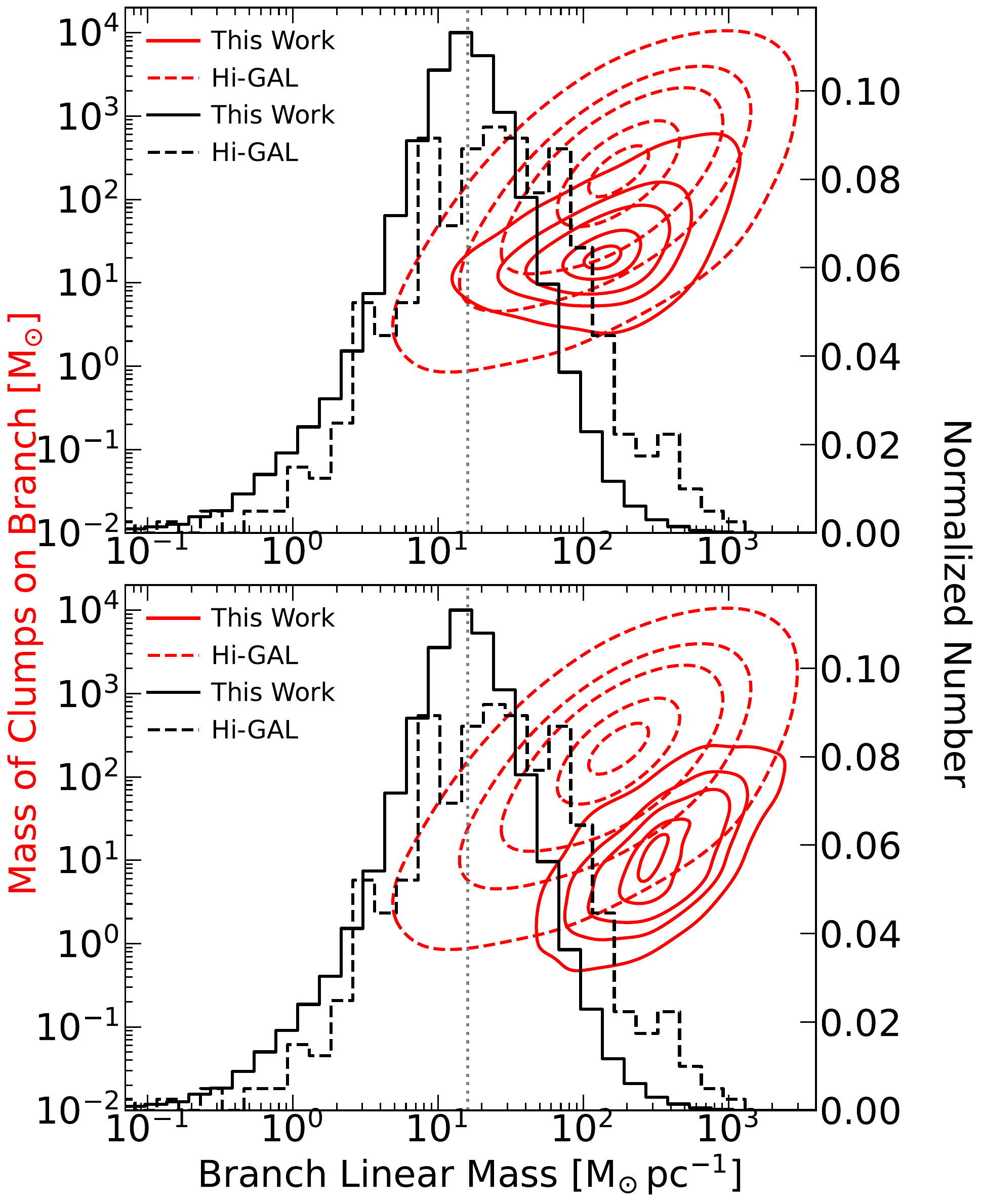}  
\caption{\textit{Top:} clump mass vs.\ host line mass ($M_{\rm line}$). Red contours (levels: $90\%$, $70\%$, $50\%$, $20\%$, and $5\%$) represent synthetic branches (solid) and Hi-GAL filaments \citep[dashed;][]{Schisano+2020MN}. Black histograms (right y-axis) show the $M_{\rm line}$ distribution for clump-free structures (solid: this work, dashed: Hi-GAL). The vertical dotted line indicates the critical line mass ($16\,\rm M_\odot\,pc^{-1}$ at $T=10\,\rm K$). \textit{Bottom:} same as the top, but for clumps with embedded stars (using 3D gas mass).}
\label{fig_app_linear_mass}  
\end{figure}

Figure~\ref{fig_app_linear_mass} also shows a similar correlation between clump mass and branch line mass. Both datasets follow similar scaling relations driven by gravitational fragmentation \citep{Andre.P+2010A&A, Arzoumanian+2011A&A}, indicating that the fragmentation mechanism in branches may be identical to that in the main filaments. 

\twocolumngrid

\bibliography{references}{}
\bibliographystyle{aasjournal}




\end{document}